\documentclass[twocolumn,showpacs,preprintnumbers,amsmath,amssymb,superscriptaddress,10pt,aps,prb]{revtex4-2}

\usepackage{graphicx}
\usepackage[colorlinks=true, allcolors=blue]{hyperref}
\usepackage{braket}
\usepackage{csquotes}
\usepackage[dvipsnames]{xcolor}
\usepackage{caption}
\usepackage{subcaption}

\begin{document}

%\preprint{APS/123-QED}

\title{Striped Spin Density Wave in a Graphene/Black Phosphorous Heterostructure}

\author{Dolev Haddad}
\affiliation{Department of Physics, Bar-Ilan University, Ramat Gan 52900, Israel}
\author{H.A. Fertig}
\affiliation{Department of Physics, Indiana University, Bloomington, IN 47405, USA}
\affiliation{Quantum Science and Engineering Center, Indiana University, Bloomington, IN 47408, USA}
\author{Efrat Shimshoni}
\affiliation{Department of Physics, Bar-Ilan University, Ramat Gan 52900, Israel}

\begin{abstract}
A bilayer formed by stacking two distinct materials creates a moiré lattice, which can serve as a platform for novel electronic phases. In this work we study a unique example of such a system: the graphene-black phosphorus heterostructure (G/BP), which has been suggested to have an intricate band structure. Most notably, the valence band hosts a quasi-one-dimensional region in the Brillouin zone of high density of states, suggesting that various many-body electronic phases are likely to emerge. We derive an effective tight-binding model that reproduces this band structure, and explore the emergent broken-symmetry phases when interactions are introduced. Employing a mean-field analysis, we find that the favored ground-state exhibits a striped spin density wave (SDW) order, characterized by either one of two-fold degenerate wave-vectors that are tunable by gating. Further exploring the phase-diagram controlled by gate voltage and the interaction strength, we find that the SDW-ordered state undergoes a metal to insulator transition via an intermediate metallic phase which supports striped SDW correlations. Possible experimental signatures are discussed, in particular a highly anisotropic dispersion of the collective excitations which should be manifested in electric and thermal transport.
\end{abstract}

\maketitle
\section{Introduction}
\label{sec:introduction}
Two-dimensional (2D) materials, such as graphene and transition metal dichalcogenides (TMDs), have emerged in  recent years as  appealing platforms for realizing and studying a rich variety of electronic phases. A prominent feature of these materials, whose structure is dictated by van der Waals forces between atomically thin layers, is their tunability by a multitude of means. Among them, hybrid materials created by stacking of two or more layers with a lattice mismatch or a twist angle, play a key role in recent novel discoveries \cite{bistritzer2011moire,gong_2014,Liu_2016,Ajayan_2016,duong2017van,cao_2018,cao2018unconventional,yankowitz2019tuning,carr2017twistronics,novoselov20162d,Lu_2019}. 
The main feature shared by this family of materials is the formation of a moiré pattern or a superlattice structure characterized by a large lattice constant, typically much larger than the atomic scale lattice constants of each of the monolayers from which it is constructed. The resulting band-structure exhibits flattening of certain minibands, leading to enhancement of correlation effects. This sets the stage for the formation of diverse interaction-driven electronic phases
\cite{andrei2021marvels,Choi2021,Wang2022,mak2022semiconductor,wietek2022tunable,zhou2023QuantumMelting,Foutty2024}.

The extensive study of such van der Waals materials was initiated by the isolation of graphene, a two-dimensional allotrope of carbon arranged in a honeycomb lattice \cite{neto2009electronic}. A prominent aspect of graphene is that its band structure features two Dirac cones, forming distinct valleys at the $K$ and  $K'$ points in the Brillouin zone.
As a result, monolayer graphene is a semimetal which manifests unique electronic properties owing to the massless Dirac fermion nature of its low-energy excitations. Stacking of additional layers of graphene or other 2D materials often generates systems with dramatically different properties than those of a single pristine graphene sheet. Moreover, their more complex structure allows for the manipulation of their ground state properties, for example by the application of interlayer displacement fields.  

\begin{figure}
    \begin{subfigure}{0.46\textwidth}
       \caption{}
        \includegraphics[scale=0.1]{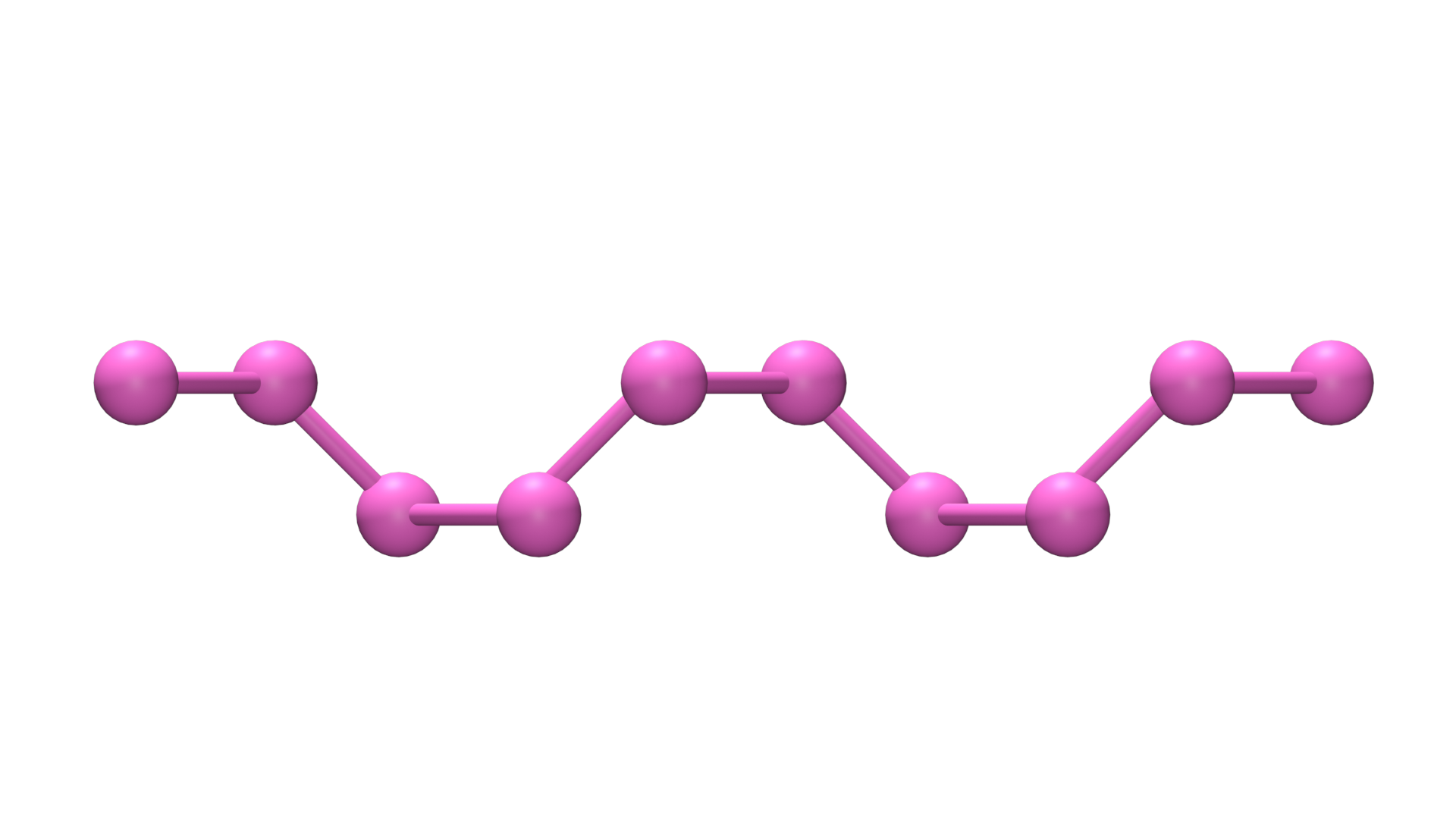}
    \end{subfigure}
    \begin{subfigure}{0.23\textwidth}
        \caption{}
        \includegraphics[scale=0.08]{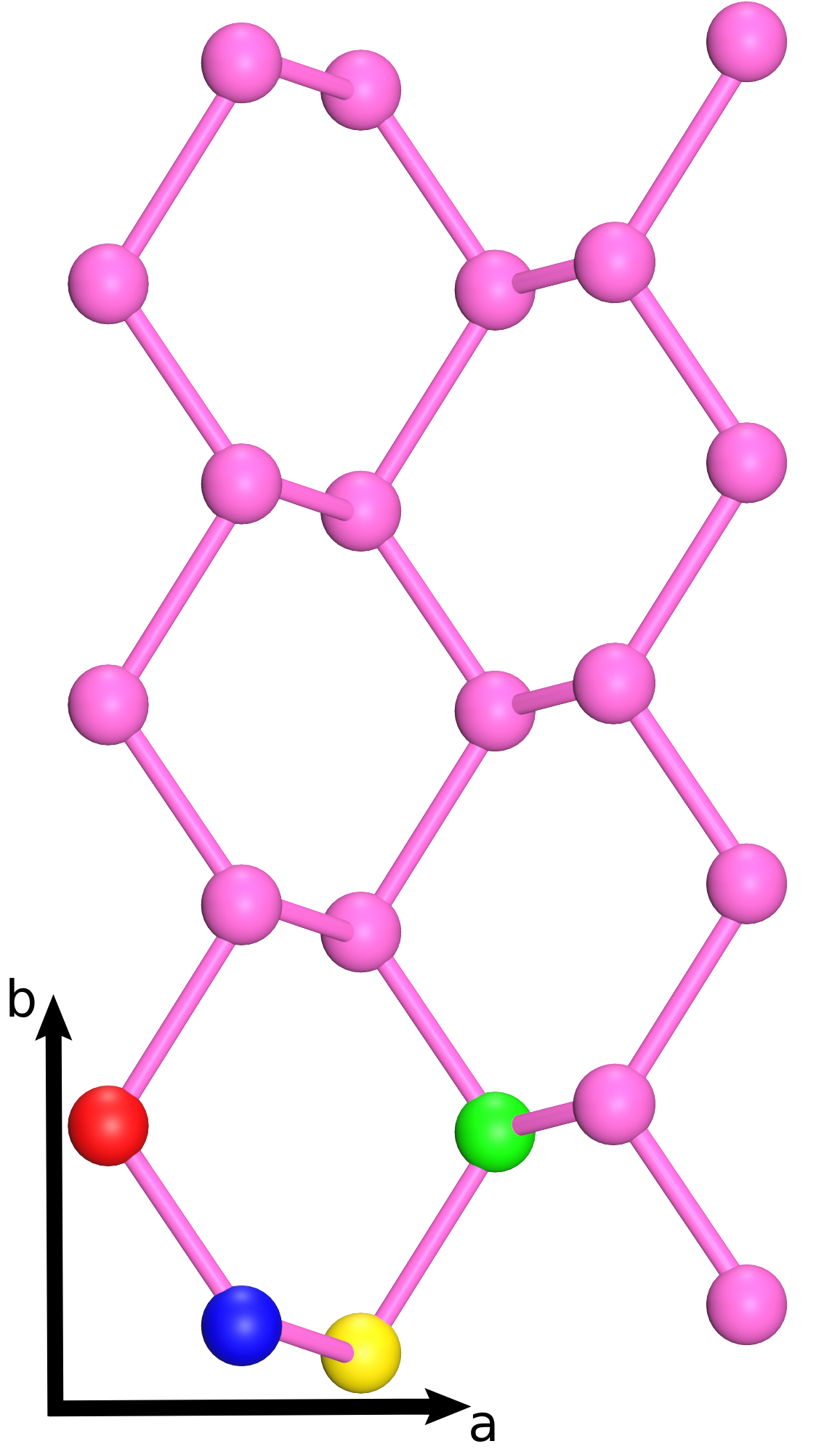}
        \label{BP_2}
    \end{subfigure}
    \begin{subfigure}{0.23\textwidth}
        \caption{}        
        \includegraphics[scale=0.2]{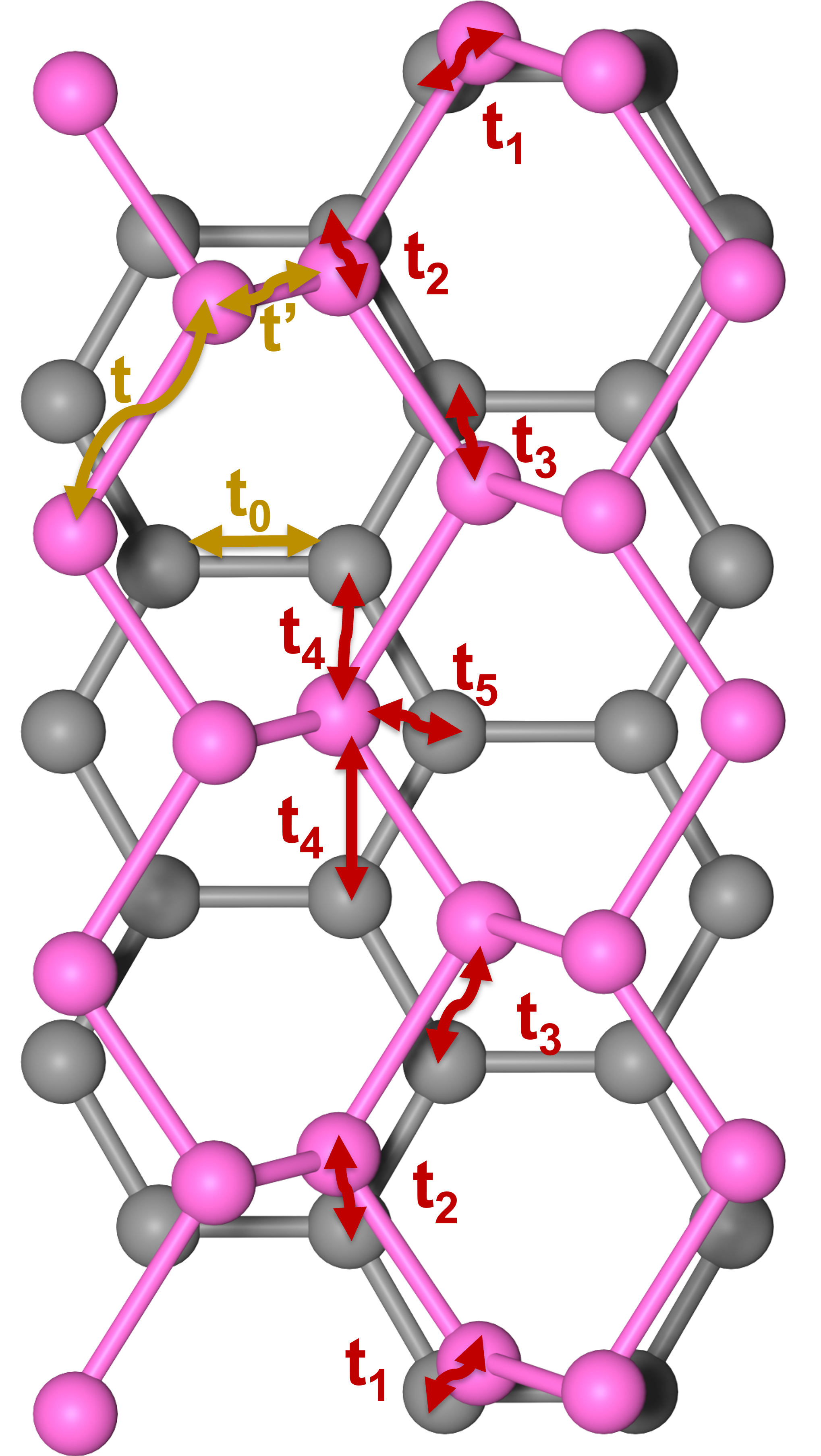}
        \label{fig:G/P unit cell}
    \end{subfigure}
    \caption{(a) Side view of the black phosphorous crystal structure. (b) Crystal structure of black phosphorous and its unit cell (colored); it contains four sites which are color-coded as $u$ for red-blue (upper layer zigzag) and $d$ (lower layer zigzag) for yellow-green. Within the $u$ ($d$) layer, the red (yellow) and blue (green) sites are labeled by $l$ (left) and $r$ (right), respectively. 
    (c) Unit cell of the graphene/phosphorene heterostructure with the hopping parameters of the tight binding. }
    \label{BP unit cell}
\end{figure}

The remarkable properties of these materials and their heterostructures has inspired the study of other types of monolayer crystals. In particular, monolayer black phosphorus (BP),  a 2D allotrope of phosphorus (also known as phosphorene) has attracted much attention. BP hosts a finite direct band gap and can be created with high mobility \cite{mu2019two,ryder2016chemically,zhao2017recent,li2014modulation,li2014black,xia2014rediscovering,koenig2014electric,valagiannopoulos2017manipulating}. It is distinct from most other 2D van der Waals materials in that it possesses a puckered honeycomb lattice structure (see Fig. \ref{BP unit cell}), which dictates anisotropy in its electrical and optical properties. 
Additionally, it implies an unusual response to strain \cite{jiang2014negative}, and consequently enhanced sensitivity to manipulation by mechanical means.

Motivated by the potential for opto-electronic applications \cite{xu2018graphene},
graphene/BP heterostructures (hereon dubbed G/BP heterostructures) have been recently investigated. The unique properties of each material separately (the semimetal nature of graphene, the anisotropy and tunability by strain of BP \cite{jiang2014negative}) suggest that together, they can form an interesting and versatile hybrid system.  In particular the electronic states can be manipulated by gating, pressure and directional strain, over a wide range of parameters.

A recent density functional theory (DFT) 
study of this system \cite{cai2015electronic} indicates the formation of a new electronic band structure with fascinating properties. Charge transfer elevates the valence band of BP closer to the Fermi level, and it strongly hybridizes with the lower Dirac cones of graphene. Most notably, the band structure of this new material hosts energy contours with long stretches of little curvature.  These regions of the Brillouin zone (BZ)
contribute to a high density of states, suggesting that various many-body electronic phases are likely to emerge.
Indeed, this is hinted at by striped patterns manifested in scanning tunneling microscopy (STM) studies of the system \cite{liu2018tailoring}. 

In this paper, we present a model for interacting electrons in a G/BP heterostructure and explore the resulting broken symmetry phase, which turns out to be a striped spin density wave (SDW). Motivated by the unique anisotropic band structure of the G/BP proposed in \cite{cai2015electronic}, we construct a tight-binding model of the G/BP heterostructure, which recovers its prominent features. We then extend our model by introducing repulsive contact interactions, and analyze it within a mean-field (MF) approximation. Specifically, we proposed two distinct order parameters, encoding spin density wave (SDW) and charge density wave (CDW) order in the ground state [see Eq. (\ref{OP}) below]. We find that SDW order is more the more energetically favorable of the two, and that it features a striped nature. Interestingly, the striped SDW order is supported in two distinct phases, a metallic and an insulating one. The resulting low-temperature phase diagram is presented in Fig. \ref{fig: Phase Diagrams}  below. 

The remainder of this paper is organized as follows.
In Section \ref{ESBH} we present our derivation of the tight-binding model for the G/BP heterostructure, and propose an effective continuum model capturing its main features. In Section \ref{MF_Analysis} we introduce the model for interactions and detail its analysis  using a mean-field approximation. The principal results of this analysis are presented
in Section \ref{Sec:results}.  
In section \ref{Discussion} we summarize our conclusions and provide some outlook. Finally, more elaborate technical details are presented in Appendices \ref{AppA} and \ref{spin_exVal}.

\section{The non-interacting Hamiltonian}\label{ESBH}  As a first step, we seek to develop a simple model that captures the main features of the band structure presented in Ref. \onlinecite{cai2015electronic}. This approach will allow us to add  interactions in a form amenable to further analysis. To achieve this, we first analyze the isolated BP structure, and then use this to construct a tight-binding model for the hybrid structure. Subsequently, we show that prominent features of the upper valence band are recovered by an effective continuum model. 
\subsection{Tight-binding model for the BP layer}\label{t.b}
The unit cell of an isolated BP layer contains four atoms, arranged within a rectangle unit cell, with primitive lattice vectors $\mathbf a = a \hat x \quad \textrm{and} \quad \mathbf b = b \hat y$.  In the unit cell, two of the atoms reside in an ``upper layer,'' and two in a ``lower layer,'' each of which is an phosphorus atom.  (See Fig. \ref{BP unit cell}(b).)
With each atom in the structure we associate an atomic orbital, represented as a ket of the form $\ket{\mathbf{R},i,j}$, where $\mathbf{R} $ denotes the location of a specific unit cell in the Bravais lattice, $ i $ takes values $u$ or $d$ and $ j $  is either $l$ or $r$  (see caption of Fig. \ref{BP unit cell}).
The BP lattice can be understood as an array of coupled zigzag chains, residing fully either in the upper and lower layers. Because of this, there are hopping parameters connecting sites in the same layer, and sites in opposite layers. Thus a minimal model involves two types of hopping parameters, which we specifiy as $ t $ for intra-layer hopping, and $ t' $ for inter-layer hopping.
This leads to a tight-binding Hamiltonian of the form
\begin{widetext}
    \begin{equation}
        \begin{split}
            H= & -t \sum_{\mathbf R}\bigg \{ \ket{\mathbf{R},u,l} \left[\bra{\mathbf{R}+\mathbf b,u,r}+\bra{\mathbf{R},u,r}\right]+ \ket{\mathbf{R},u,r} \left[\bra{\mathbf{R} - \mathbf b,u,l}+\bra{\mathbf R,u,l}\right]
            \\&+ \ket{\mathbf{R},d,l} \left[\bra{\mathbf R - \mathbf b,d,r}+\bra{\mathbf R,d,r }\right] + \ket{\mathbf{R},d,r}\left[\bra{\mathbf R+\mathbf b,d,l}+\bra{\mathbf R,d,l}\right] \bigg \}
            \\& -t'\sum_{\mathbf R}\left \{ \ket{\mathbf R,u,l} \bra{\mathbf R- \mathbf a,d,r}+ \ket{\mathbf{R},u,r} \bra{\mathbf R,d,l} +\ket{\mathbf{R},d,l} \bra{\mathbf R,u,r}+\ket{\mathbf{R},d,r} \bra{\mathbf R+ \mathbf a,u,l} \right \}.
        \end{split}
    \end{equation}
\end{widetext}
Rewriting this in the plane wave basis, $\lvert{\mathbf{k}},i,j\rangle \equiv \frac{1}{\sqrt{N}} \sum_{{\mathbf R}} \lvert{\mathbf{R}},i,j\rangle$, with $N$ the number of unit cells, the Hamiltonian takes the form
$H=\sum_{\mathbf{k}} \psi_{\mathbf{k}}^\dagger \hat H_{\mathbf{k}} \psi_{\mathbf{k}}$,
where we have introduced the 4-spinors
\begin{equation}\label{key3}
	\psi_{\mathbf k}^\dagger|0\rangle=\big( \lvert \mathbf{k},u,l \rangle, \lvert \mathbf{k},u,r \rangle, \lvert \mathbf{k},d,l \rangle, \lvert \mathbf{k},d,r \rangle \big)
 \quad
\end{equation}
and the $\mathbf{k}$-dependent Hamiltonian
\begin{equation}\label{key4}
\resizebox{0.45\textwidth}{!}{%
 	$\hat{H}_{\mathbf{k}}=\begin{pmatrix}
 	0 & -t(e^{i \mathbf{k} \cdot \mathbf{b}}+1) & 0 & -t' e^{-i \mathbf{k} \cdot \mathbf{a}} \\
 	-t(e^{-i \mathbf{k} \cdot \mathbf{b}}+1) & 0 & -t' & 0 \\
 	0 & -t' & 0 & -t_2(e^{-i \mathbf{k} \cdot \mathbf{b}}+1) \\
 	-t' e^{i \mathbf{k} \cdot \mathbf{a}} & 0 & -t(e^{i \mathbf{k} \cdot \mathbf{b}}+1) & 0
\end{pmatrix}$}.
\end{equation}

This yields the 4-band energy spectrum
\begin{equation}\label{BP-energy}
	\epsilon^{c/v}_{\pm,\mathbf{k}}= \pm \bigg [ 4t^2 \cos^2(\frac{k_y b}{2})+t'^2 \pm 4tt' |\cos(\frac{k_y b}{2})\cos(\frac{k_x a}{2})|\bigg ]^{1/2},
\end{equation}
where $c/v$ stands for conduction and valence bands, which are positive and negative, respectively. For an estimate, we may consider the hopping parameter values for an isolated BP layer, \cite{sisakht2015scaling}  $t \approx -1.2\,\text{eV}$ and $t' \approx 3.6\,\text{eV}$, yielding a band gap of $\Delta \varepsilon = 2|2t+t'| \approx 2.4\,\text{eV}$. However, in a heterostructure environment, $t$ and $t'$ may be significantly modified, so we regard them as free parameters.

\subsection{Tight-binding model of the G/BP heterostructure}\label{ssec: T.B.}

We next construct a tight-binding model for the coupled G/BP system, with the graphene layer below the BP layer.  We assume direct tunneling between these layers only to the bottom zig-zag chain of BP.

Fig. \ref{fig:G/P unit cell} presents the full unit cell we use for the heterostructure, in which we have assumed a small distortion in the BP unit cell. 
The Bravais lattice vectors of the supercell are assumed to be $\mathbf{a_1}=3a_0\hat x, \; \mathbf{a_2}=4\sqrt{3}a_0\hat y$, where $a_0=1.42 \text{\AA}$ is the graphene lattice constant. 
The resulting commensurate structure has a supercell containing 16 carbon atoms in graphene layer and 12 atoms in the BP layer, which closely resembles the structure considered in \cite{cai2015electronic}.  We introduce hopping parameters between the two layers, as illustrated in Fig. \ref{BP unit cell}
($t_0$ for nearest neighbor carbon atoms in the graphene layer; $t$ and $t'$ in the BP layer; and $t_1-t_5$ for inter-layer tunneling). Among the inter-layer hopping parameters, we expect $t_{1} \approx t_{2}$, both corresponding to nearly vertically aligned carbon-phosphorus sites, will be larger than the other interlayer hopping parameters. Finally, we introduce an on-site energy $ \epsilon_{G} $ for the graphene sites relative to the BP.
Note that, in principle, the tunneling parameters can be modified by applying pressure to the system.  Moreover, on-site energies may be modified by application of an electric field.

\begin{figure}[h]
    \begin{subfigure}{0.45\textwidth}
        \caption{}
        \includegraphics[width=0.9\textwidth]{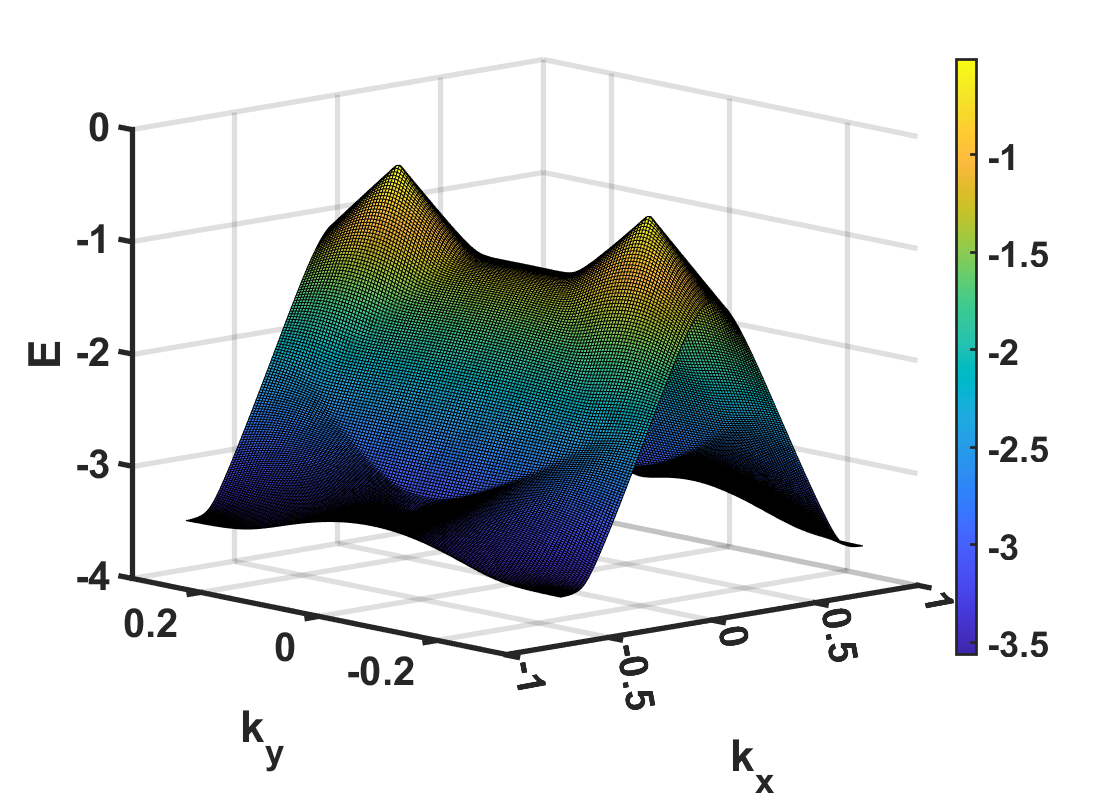}
    \end{subfigure}
    \begin{subfigure}{0.45\textwidth}
        \caption{}
        \includegraphics[width=0.9\textwidth]{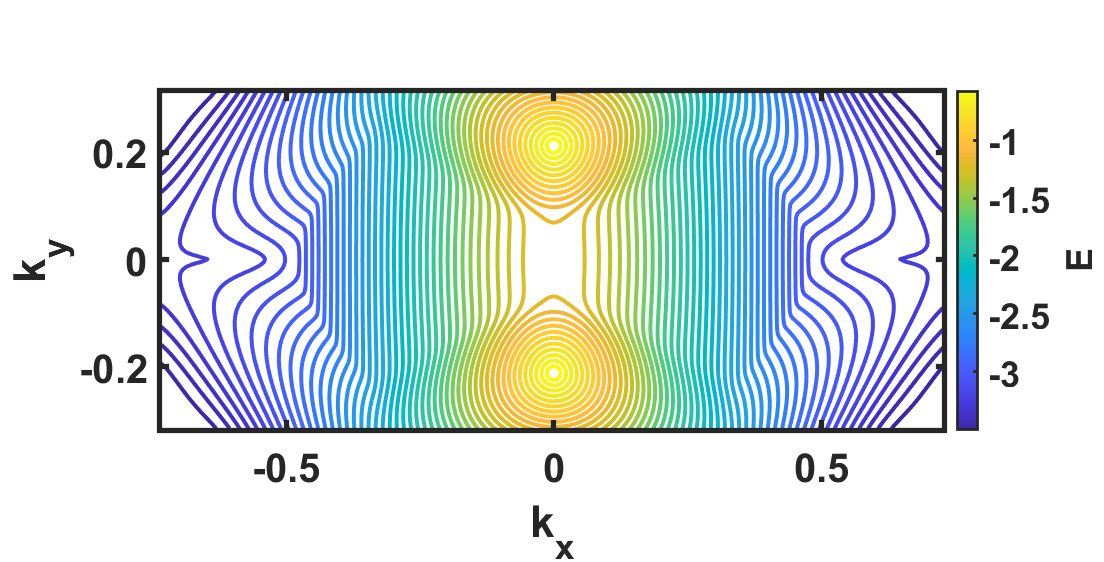}
        \label{tight binding: b}
    \end{subfigure}
    \caption{Tight-binding energy dispersion of the top valance band of the G/BP heterostructure. Here $t_1=t_2=0.35 \,\text{eV}$, $t_6=t_8=0.15\,\text{eV}$, $t_7=0.1\,\text{eV}$ and assuming an on-site energy on the graphene layer of $\epsilon_G=-0.5\,\text{eV}$. (a) 3D plot; here $E$ is in units of $\text{eV}$, and $\mathbf{k}$ in $(\textup{~\AA})^{-1} $. (b) Energy contours of the corresponding valance band.}
    \label{tight binding}
\end{figure}

\begin{figure}[h]
    \includegraphics[scale=0.29]{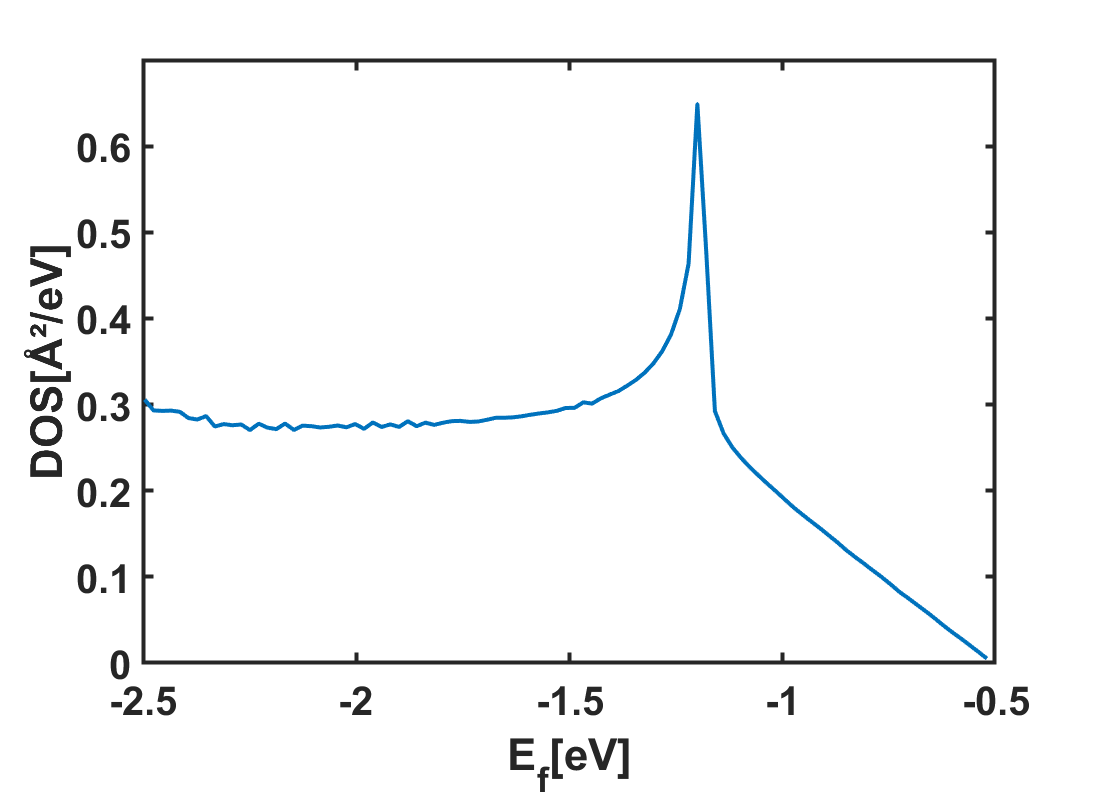}
    \caption{Numerically generated DOS corresponding to the band structure of Fig. \ref{tight binding} for energies in the uppermost valence band.}
    \label{DOS}
\end{figure}

In Fig. \ref{tight binding} we present the highest filled (valence) band resulting from a numerical calculation of the tight-binding spectrum. This band is readily accessible by moderate hole doping, and has interesting properties, the consequences of which are the focus of our study. We thus refer to it in what follows as the active band.
One special characteristic of this band is that near the supercell $\Gamma$ point there is an extended, nearly dispersionless region, below which (in energy) are energy contours with very weak  $k_y$ dependence (see Fig. \ref{tight binding}(b).) These features are also reflected in the density of states (DOS) presented in Fig. \ref{DOS}, which has a sharp peak at $E \sim -1.2\,\text{eV}$  corresponding to the energy at the $\Gamma$ point, and a wide plateau below this.

Setting the Fermi energy to a value in the plateau region of the DOS (i.e. within the range $E_F\in [-2.5\,\text{eV}, -1.5\,\text{eV}]$), the Fermi surface supports a pair of nearly straight, parallel lines separated by a well-defined wave-vector $\mathbf Q_0 = 2k_F \hat x$. This nesting structure makes this system a strong candidate for density-wave instabilities, as we will show below.  In order to explore the origin of this flattening in the $k_y$-dispersion of the active band, we next develop a simplified continuum Hamiltonian that focuses on the band structure for energies in this plateau region.

\subsection{\label{3 band} Effective continuum model}
\begin{figure}
	\centering
	\includegraphics[width=0.47\textwidth]{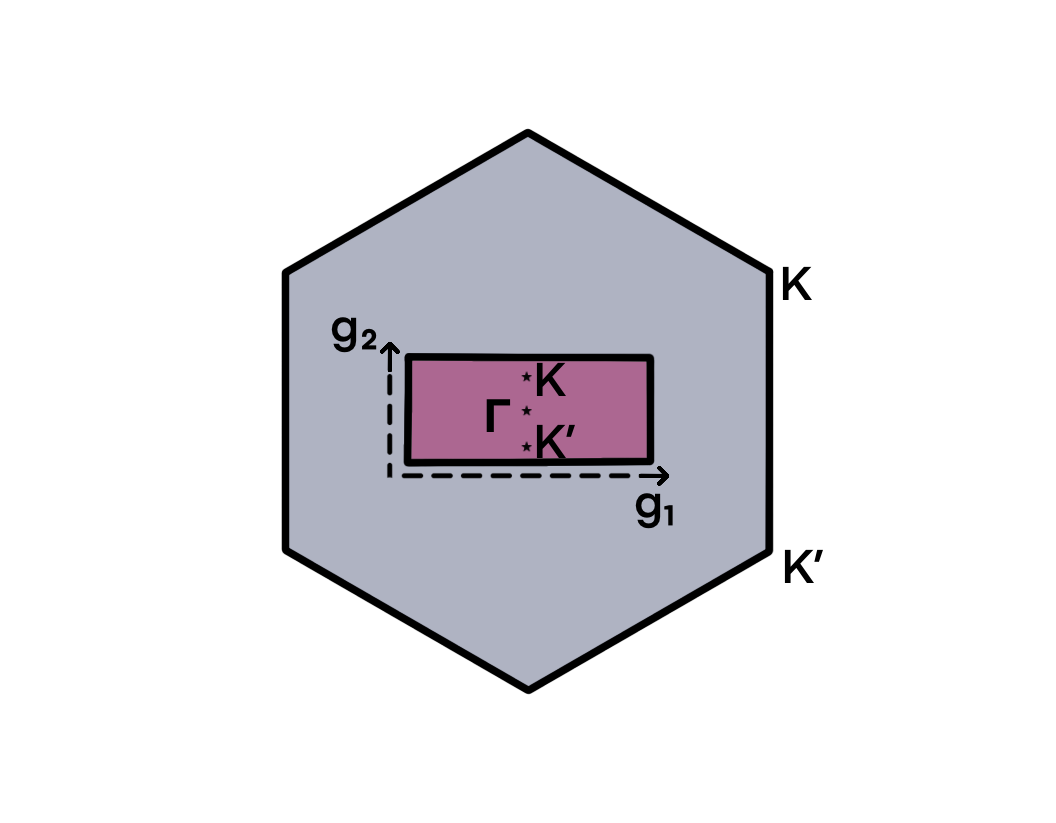}
	\caption{Brillouin zone (BZ) of the G/BP heterostructure (purple rectangle) inside the graphene BZ. Here $ \mathbf{g_{1}}, \ \mathbf{g_{2}}$ are the reciprocal lattice vectors of the super-cell. Each valley is mapped to a single point along the y axis in the small BZ; these correspond to the two Dirac points observed in Fig. \ref{tight binding}.}
 \label{foldedBZ}
\end{figure}

We next derive a simplified three-band Hamiltonian which captures the important features of the tight-binding band structure in the energy range of interest.
As a first step we consider how the graphene and BP Brillouin zones fold into the first BZ of the  superlattice, which appears as a rectangle in Fig. \ref{foldedBZ}. For the commensurate structure we consider, the superlattice reciprocal lattice vectors are
\begin{equation}\label{hetero-rl}
\mathbf{g_1} = (\frac{2 \pi}{3a_{0}},0), \; \mathbf{g_2} = (0,\frac{\pi}{2 \sqrt{3}a_{0}}).
\end{equation}
After zone-folding, the graphene $\mathbf{K}$ point is mapped to $(0, \frac{\pi}{6 \sqrt{3}a_{0}})$ inside the superlattice BZ, while the $ \mathbf{K}' $ point maps to $ (0, \frac{-\pi}{6 \sqrt{3} a_{0}}) $.  This is illustrated in Fig. \ref{foldedBZ}.

Remarkably, the main features of the commensurate tight-binding system described in the previous section can be captured by a three-band model with a single phenomenological hybridization variable.  The model is defined by starting with the bare graphene dispersion,
\begin{equation}\label{key8}
	\epsilon_{K,K'} = -t_{0}|f_{\pm}(\mathbf{k})|+\epsilon_{G}
\end{equation}
with $+/-$ referring to the $K/K'$ point, respectively; $t_0\approx 2.8 \, \text{eV}$ \cite{neto2009electronic}, and

\begin{equation}
    \begin{split}
        & f_{\pm}(\mathbf{k}) = 1+2\cos(\frac{\sqrt{3}a_{0}}{2} q^{\pm}_{y}) e^{\frac{i3a_{0}}{2} q^{\pm}_{x}}\, ,\\& {\rm where}\quad\mathbf{q}^{\pm} =\mathbf{k} + \mathbf{g}_{1} \pm \mathbf{g}_{2}\, .
    \end{split}
\end{equation}
Here $\mathbf{k}$ is a point in the superlattice BZ, and $\mathbf{q}^{\pm}$ is the corresponding point in the pristine graphene BZ.  Note the parameter $\epsilon_G$ introduces an offset of the graphene band structure, as a result of being brought into the proximity of the BP layer.

\begin{figure}
    \centering
    \begin{subfigure}{0.23\textwidth}
        \centering
        \caption{}
        \includegraphics[width=1\textwidth]{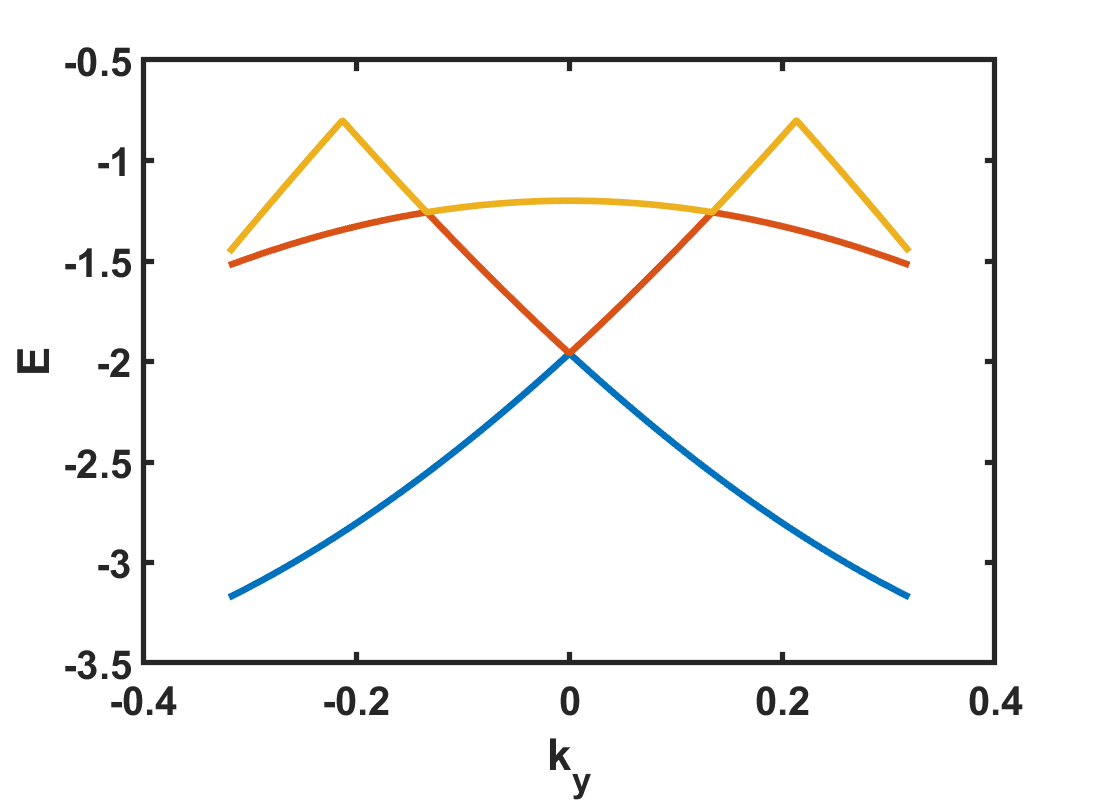}
        \label{t4}
    \end{subfigure}
    \begin{subfigure}{0.23\textwidth}
        \centering
        \caption{}
        \includegraphics[width=1\textwidth]{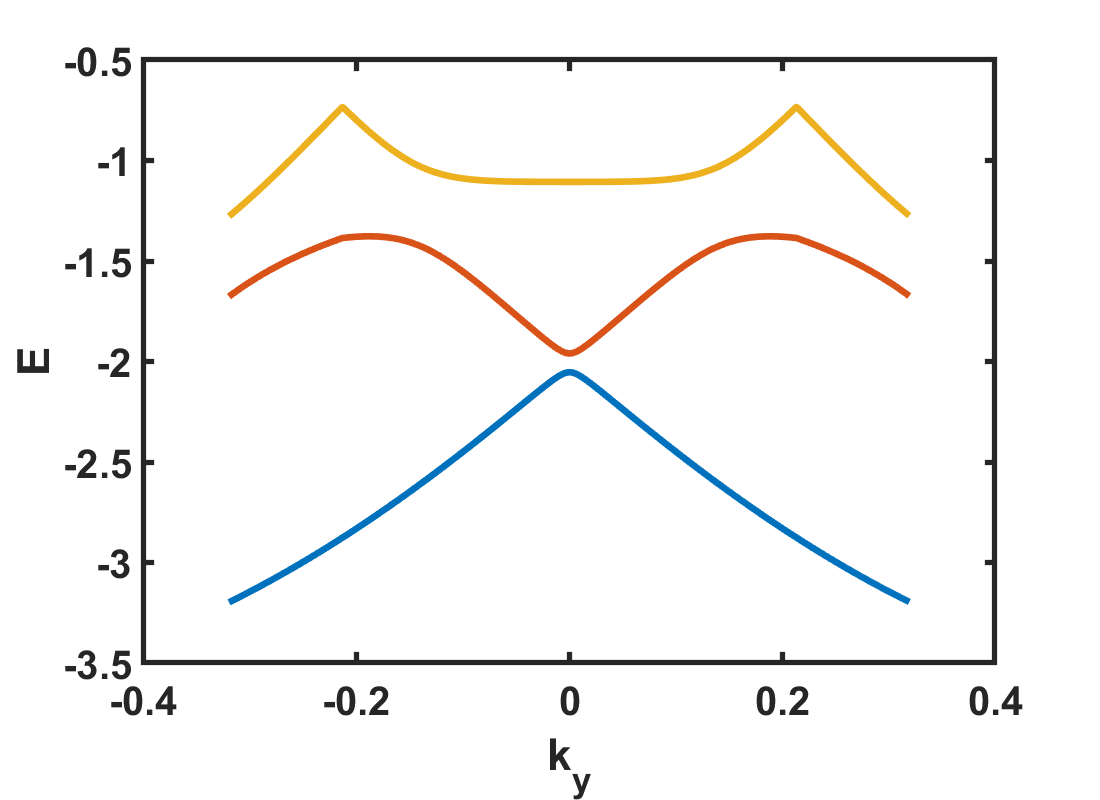}
        \label{Heff}
    \end{subfigure}
    \begin{subfigure}{0.23\textwidth}
        \centering
        \caption{}
        \includegraphics[width=1\textwidth]{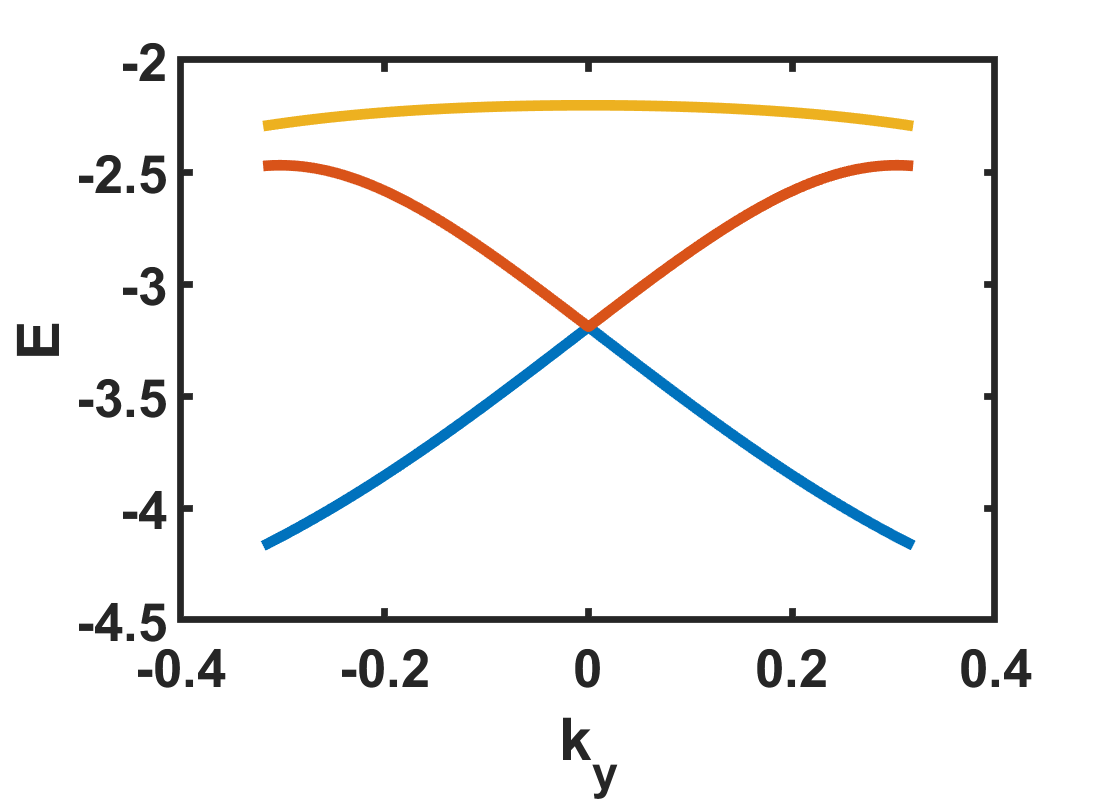}
        \label{sfig: contNo_t4}
    \end{subfigure}
    \begin{subfigure}{0.23\textwidth}
        \centering
        \caption{}
        \includegraphics[width=1\textwidth]{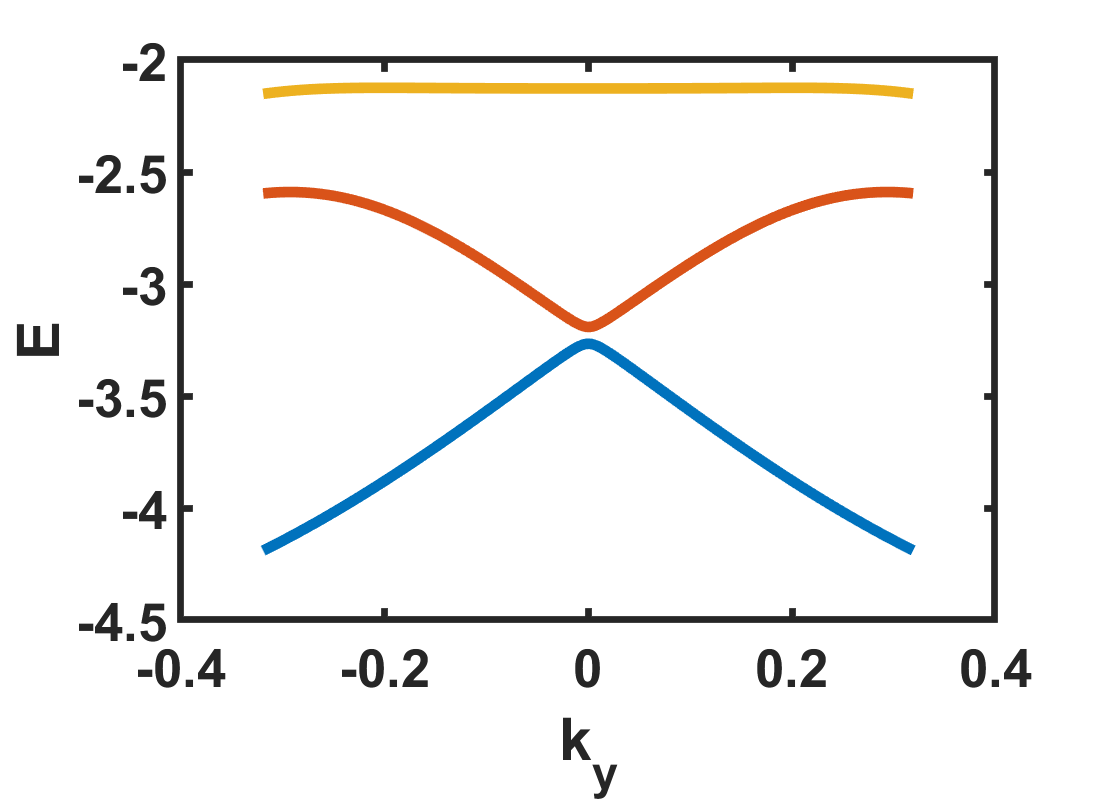}
        \label{sfig: cont}
    \end{subfigure}
    \begin{subfigure}{0.46\textwidth}
        \centering
        \caption{}
        \includegraphics[width=1\linewidth]{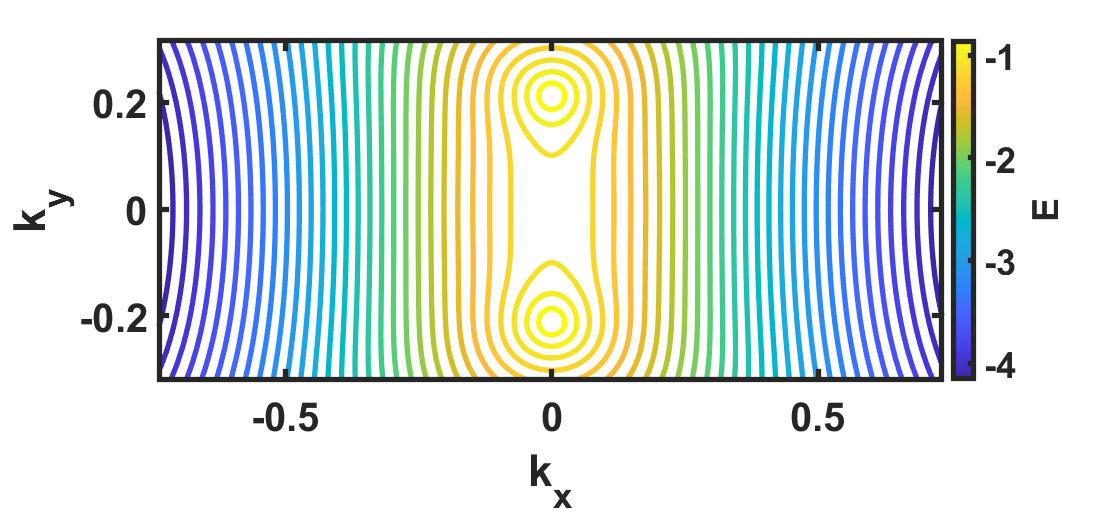}
        \label{fig: Effective Hamiltonian contour plot}
    \end{subfigure}
\caption{Cross-sections of the band structure resulting from the effective continuum model for the known parameters $t$, $t'$, $t_0$ (see text) and $\epsilon_G=-0.8 \,\text{eV}$. Top: for $k_x=0$ and
inter-layer hopping parameter (a) $t_1=0 $, (b) $t_1=0.2 \, \text{eV}$. Middle: $k_x=0.3$ and (c) $t_1=0$, (d) $t_1=0.2 \,\text{eV}$. (e) Contour plot of the active (topmost) band derived from the effective Hamiltonian Eq. (\ref{eq:Heff}) for the same parameters and  
$t_1=0.2\, \text{eV}$.
In comparison with Fig. \ref{tight binding}, the effective continuum model retains the key features: the dispersionless region near the $\Gamma$ point, and below that, the quasi-one dimensional energy contours.}
    \label{crossection}
\end{figure}

In terms of these, our minimal model for the Hamiltonian of the non-interacting G/BP heterostructure has the form $ H_{eff} = \sum_{\mathbf{k}} H_{eff}(\mathbf{k})$, with
\begin{equation}\label{eq:Heff}
	H_{eff}(\mathbf{k})=\begin{pmatrix}
		\epsilon_{K}(\mathbf{k}) & t_{1}  & 0 \\
		t_{1} & \epsilon_{BP}(\mathbf{k}) & t_{1} \\
		0 & t_{1} & \epsilon_{K'}(\mathbf{k})
	\end{pmatrix}.
\end{equation}
In this expression, $ \epsilon_{BP}(\mathbf{k}) $ is the dispersion of the topmost valence band of the BP, explicitly given by $ \epsilon_{+,\mathbf{k}}^{v} $ in Eq. (\ref{BP-energy}), and $t_1$ is the hybridization parameter.  Figs. \ref{crossection} (a),(c) illustrate how the three bands come together for $t_1=0$; Figs. \ref{crossection} (b),(d) illustrate the effect of introducing a non-vanishing $t_1$.
The topmost band resulting from this model is qualitatively quite similar to the top-most valence band of the more elaborate tight-binding model.  This model shows that the formation of the flattened region near the $\Gamma$ point of the active band arises due to hybridization of the graphene Dirac cones with the BP valence band. The dispersion of the band energy in the small $k_y$ limit is slower than quadratic; its best fit by single power-law is $\epsilon \sim k_y^4$. This behavior holds surprisingly far into the BZ, where the constant energy surfaces host long regions with little $k_y$ dependence. Indeed, when the Fermi energy is in these regions, a useful approximation for $\epsilon(\mathbf k)$ can be written in terms of  two Fermi points $(k_x,k_y)=(\pm k_F,0)$, as
\begin{equation}
    \epsilon_\pm(\mathbf k)\approx\mp v(k_x \mp k_F)+a k_y^\alpha,
    \label{toy_dispersion}
\end{equation}
where $\alpha\approx 4$.
The form of these dispersions show that the system bears a strong resemblance to a series of one-dimensional systems labeled by $k_y$. In what follows we discuss how this analogy is exploited to formulate an interaction Hamiltonian for the system.

\section{Interacting model and Mean Field Analysis}\label{MF_Analysis}

\subsection{Interaction Hamiltonian}\label{Interacting Hamiltonian}
We next add interaction terms, with an eye to investigating possible instabilities to states of broken symmetry. The most general Hamiltonian with a two-body interaction is given by
\begin{equation}
    \begin{split}
        H = & \sum_{\mathbf k, \sigma} (\varepsilon_{\mathbf k,\sigma}-E_F)c^\dagger_{\mathbf k,\sigma}c_{\mathbf k,\sigma}
        \\& + \sum_{\mathbf k,\mathbf k',\mathbf q}\sum_{\sigma,\sigma'}V(\mathbf q)c^\dag_{\mathbf k-\mathbf q,\sigma} c^\dag_{\mathbf k'+\mathbf q,\sigma'} c_{\mathbf k',\sigma'} c_{k,\sigma}    
    \end{split}
    \label{H_general}
\end{equation}
where $c_{\mathbf k, \sigma}$ are annihilation operators of electrons with lattice momentum $\mathbf k$ and spin $\sigma$, $\varepsilon_{\mathbf k,\sigma}$ are the single body energies derived from the tight binding model presented above (Section \ref{ssec: T.B.}) , and $V (\mathbf q)$ is a Fourier component of a pair interaction potential $V (\mathbf r)$. 
Noting the analogy between the form of the non-interacting Hamiltonian of this system and those of one-dimensional systems \cite{giamarchi2003quantum}, we simplify the interaction to focus on low energy processes among the electrons.
The most important of these involve scattering wavevectors $\mathbf q \sim 0$ and $\mathbf q \sim \mathbf Q$, where $\mathbf Q$ is a nesting vector, which is chosen so that translation of some portion of the Fermi surface by $\pm \mathbf Q$ nearly overlays it with some other portion of the Fermi surface.
Referring to Fig. \ref{tight binding}, one sees relatively large regions for which such nesting is possible in the energy range $E_F\in [-2.5, -1.5]$, in which a single $\mathbf Q$ connects long, straight Fermi surface contours on opposite sides of the BZ.

\begin{figure}
    \centering
    \begin{subfigure}{0.22\textwidth}
        \centering
        \caption{}
        \includegraphics[width=1\textwidth]{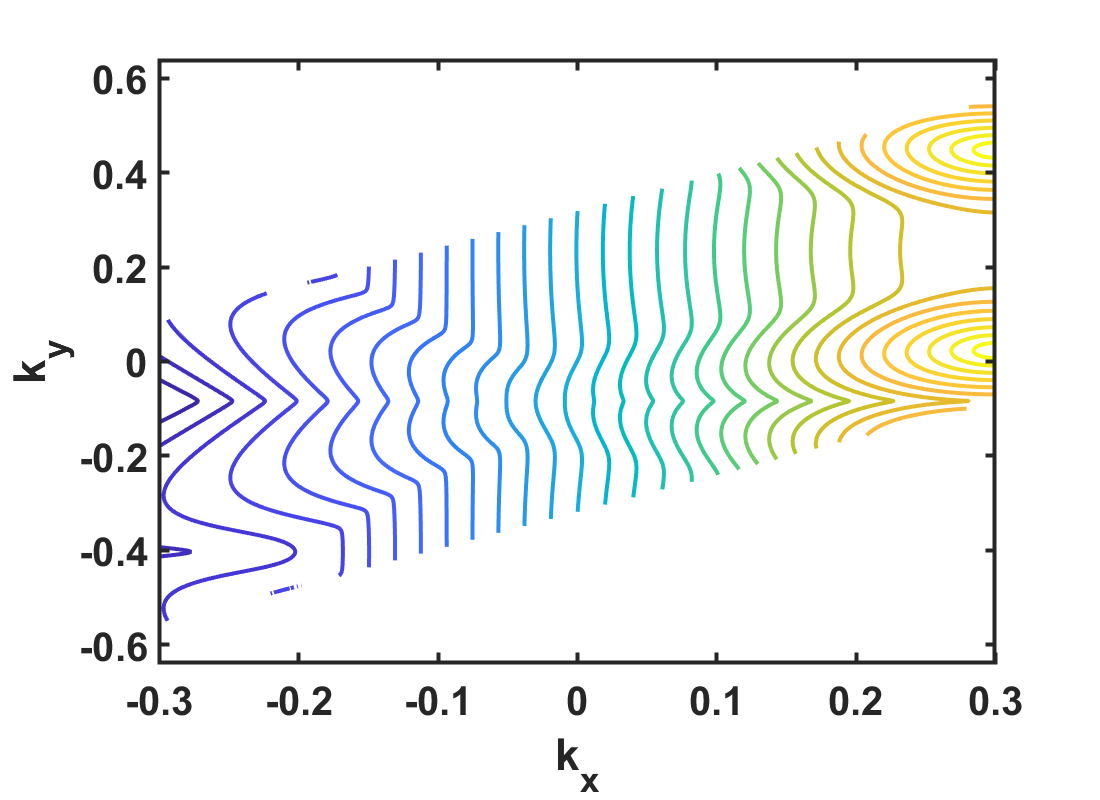}
        \label{fig:left-movers}
    \end{subfigure}
    \begin{subfigure}{0.22\textwidth}
        \centering
        \caption{}
        \includegraphics[width=1\textwidth]{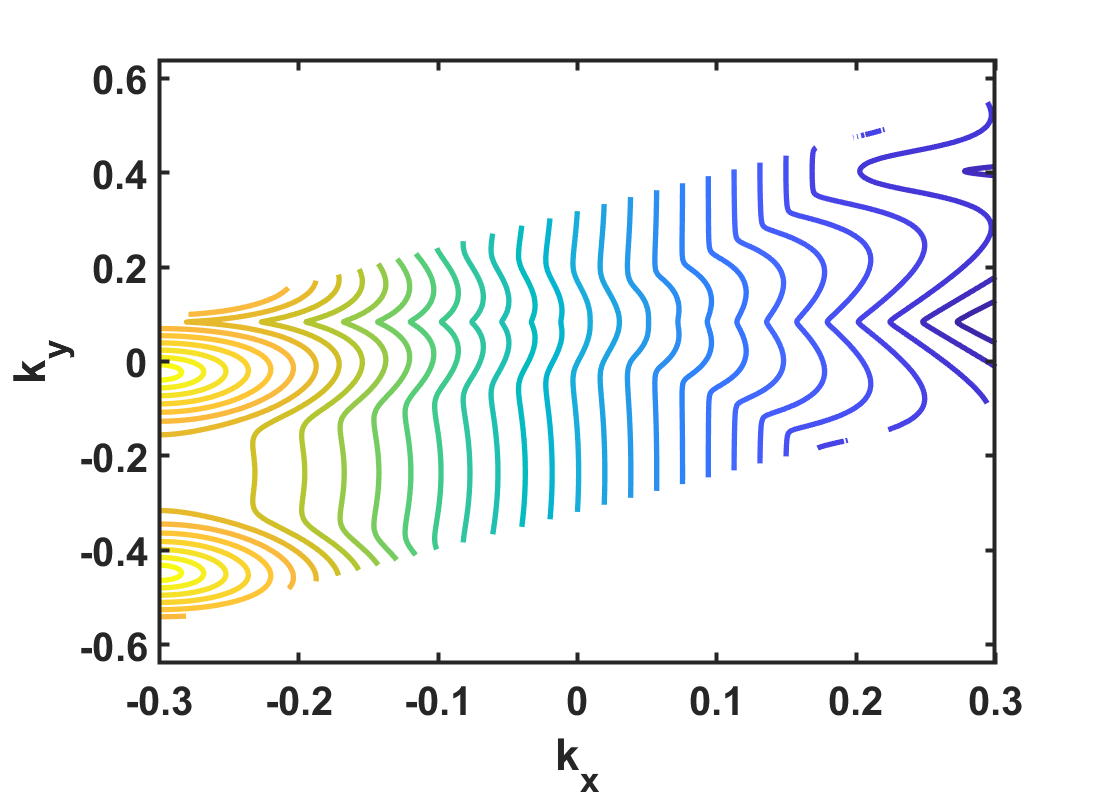}
        \label{fig:right-movers}
    \end{subfigure}
    \begin{subfigure}{0.45\textwidth}
        \centering
        \caption{}
        \includegraphics[width=1\textwidth]{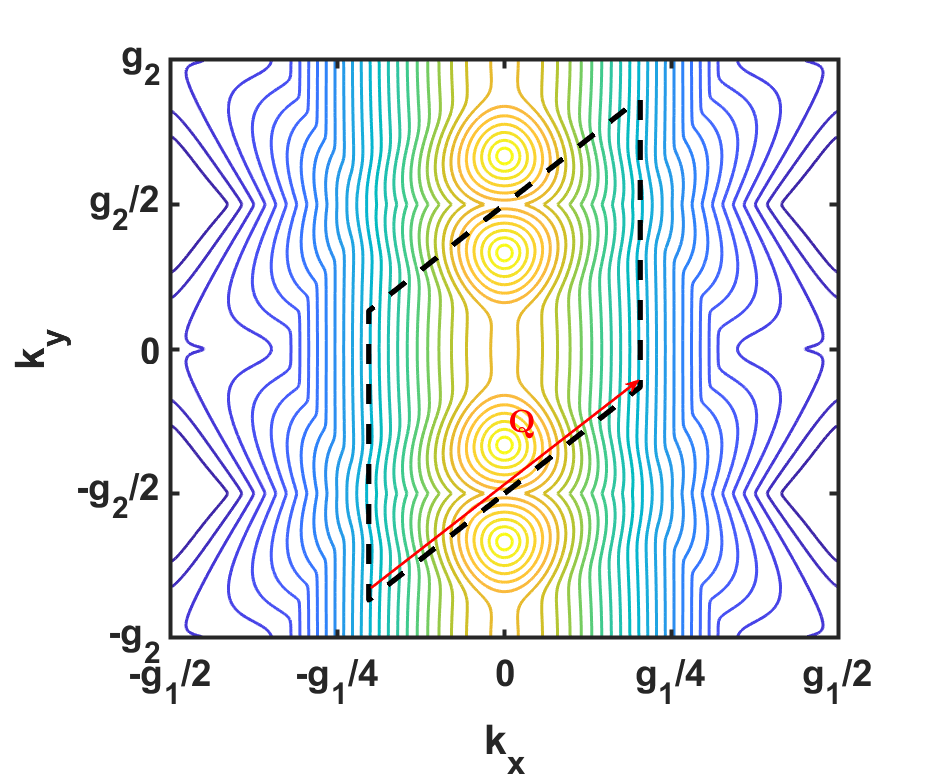}
        \label{rec_BZ}
    \end{subfigure}
    \caption{The reconstructed BZ Here $E_F=-2.2\,\text{eV}$, $\frac{1}{2}\mathbf Q =0.3 \hat x +0.235\hat y$. (a) left movers (b) right movers. (c) The reduced BZ for the same choice of $\mathbf{Q}$. 
    }
    \label{fig:reconstructedBZ}
\end{figure}

Because of this structure, 
we divide our band into two ``valleys,'' regions of the BZ near the left (L) or right (R) nearly straight Fermi surface contour. We then write our electron operators in the form $C_{\mathbf k, \tau, \sigma} \equiv c_{\mathbf k + \frac12 \tau \mathbf Q,\sigma}$ where $\tau=R,L=+,-$ denotes the Fermi line at $\frac12 Q_x=\tau k_F$, and $\mathbf k$ is defined with respect to the centers of the reconstructed BZs depicted in Figs. \ref{fig:left-movers},\ref{fig:right-movers}. For each valley $\tau$, this corresponds to a region in the original BZ shifted by $\frac12\tau\mathbf Q$ from the $\Gamma$ point (see Fig. \ref{fig:reconstructedBZ}). We note that due to the small but finite curvature of the contours, optimal nesting does not necessarily occur for $\mathbf Q=\mathbf Q_0$ parallel to the $\hat x$-axis. We therefore assume $\mathbf Q = Q_x \hat x+ Q_y \hat y$ where,  
in the numerical analysis described below, $Q_y$ serves as a variational parameter adjusted to minimize the ground state energy for a given $E_F$ (with $Q_x$ being implied by the corresponding horizontal distance between Fermi curves).

\begin{figure}[ht]
     \centering
     \begin{subfigure}[b]{0.2\textwidth}
         \centering
         \caption{}
         \includegraphics[width=1\textwidth]{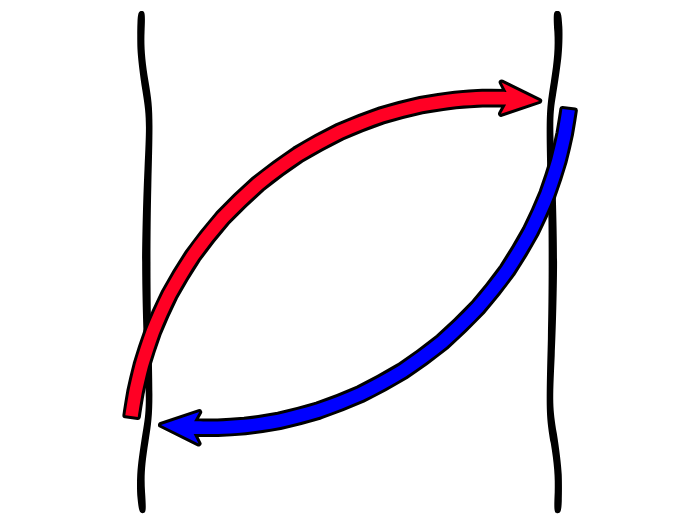}
         \label{g1}
     \end{subfigure}
     \begin{subfigure}[b]{0.2\textwidth}
         \centering
         \caption{}
         \includegraphics[width=1\textwidth]{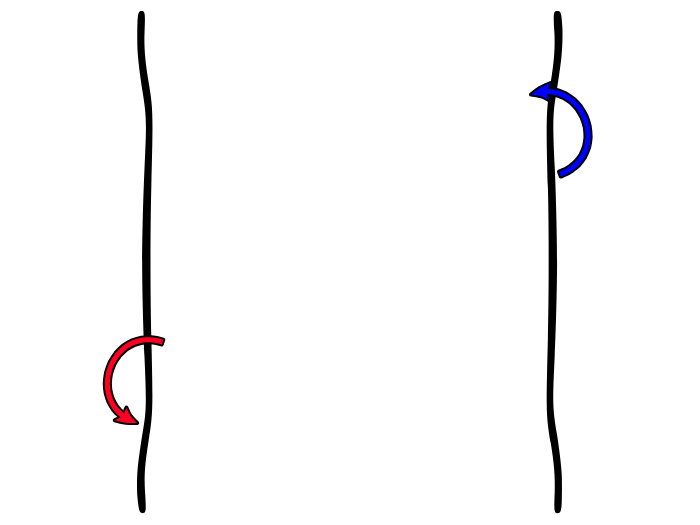}
         \label{g2}
     \end{subfigure}
		\begin{subfigure}[b]{0.2\textwidth}
         \centering
         \caption{}
         \includegraphics[width=1\textwidth]{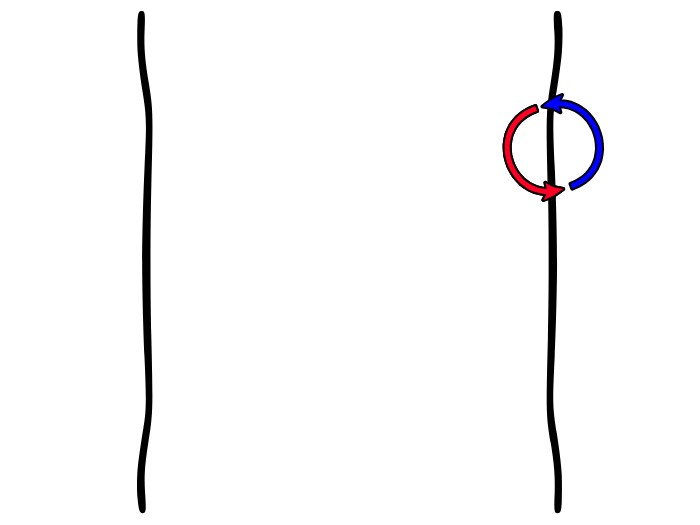}
         \label{g4}
     \end{subfigure}
        \caption{Pictorial representation of interaction terms, corresponding to the coupling constants (a) $g_1$ (b) $g_2$ and (c) $g_4$.  See text.}
        \label{interaction}
\end{figure}
We can now distinguish three kinds of interaction terms with distinct coupling parameters, summarized pictorially in Fig. \ref{interaction}.
The operators associated with these interactions are written as
\begin{equation}\label{Hint}
    \begin{split}
        & H_{g_1}=\frac{1}{2} g_1 \sum_{\substack{\tau,\sigma,\sigma' \\ \mathbf k,\mathbf k',\mathbf q}}
        C_{\mathbf k-\mathbf q,\tau,\sigma} ^\dag C_{\mathbf k'+\mathbf q,-\tau,\sigma'}^\dag C_{\mathbf k',\tau,\sigma'} C_{\mathbf k,-\tau,\sigma},
        \\& H_{g_2}=\frac{1}{2}g_2\sum_{\substack{\tau,\sigma,\sigma' \\ \mathbf k,\mathbf k',q}}
        C_{\mathbf k-\mathbf q,\tau,\sigma} ^\dag C_{\mathbf k'+\mathbf q,-\tau,\sigma'}^\dag C_{\mathbf k',-\tau,\sigma'} C_{\mathbf k,\tau,\sigma},
        \\& H_{g_4}=\frac{1}{2}g_4\sum_{\substack{\tau,\sigma,\sigma' \\ \mathbf k,\mathbf k',\mathbf q}}
        C_{\mathbf k-\mathbf q,\tau,\sigma} ^\dag C_{\mathbf k'+\mathbf q,\tau,\sigma'}^\dag C_{\mathbf k',\tau,\sigma'} C_{\mathbf k,\tau,\sigma}\, . 
    \end{split}
\end{equation}
Each of these interactions is characterized by a coupling constant, $g_i$, which is effectively a contact interaction.
$H_{g_1} $ represents an Umklapp process, involving scattering between different valleys [Fig. \ref{g1}].  $H_{g_2}$ encodes an inter-valley interaction in which particles remain in their initial valley [Fig. \ref{g2}]. Finally, $H_{g_4}$ represents an intra-valley density-density interaction, of which there is one for each valley [Fig. \ref{g4}]. For a generic repulsive interaction, all $g_i$’s are positive and $ g_1 \sim V(\mathbf{Q})<g_2, g_4$ \cite{giamarchi2003quantum}.

With these definitions, the effective Hamiltonian  for our system takes the form
\begin{equation} 
    \hat H =H_{0}+H_{g_1}+H_{g_2}+H_{g_4}
\end{equation}
in which
$$H_0=\sum_{\mathbf k,\tau,\sigma} \xi_{\mathbf k,\tau,\sigma} C_{\mathbf k,\tau,\sigma}^\dag C_{\mathbf k,\tau,\sigma}\,;$$
here 
\begin{equation} 
\xi_{\mathbf k,\tau,\sigma} = \epsilon_{\mathbf k, \tau,\sigma}-E_F\, ,\quad \epsilon_{\mathbf k, \tau,\sigma}\equiv \varepsilon_{\mathbf k+\frac12 \tau \mathbf Q,\sigma},
\label{eq:xi_def}
\end{equation}
where $\varepsilon_{\mathbf k+\frac12 \tau \mathbf Q,\sigma}$ are the single-body energies of Eq. (\ref{H_general}) near valley $\tau$. In the absence of a magnetic field imposing a Zeeman term, these energies are spin-independent allowing the definition $\xi_{\mathbf k,\tau}\equiv \xi_{\mathbf k,\tau,\sigma}$ for either $\sigma$.  

\subsection{Mean Field}
\label{MFA}

To identify broken symmetry states of this system, we next
perform a Hartree-Fock (HF) decomposition on the interaction terms.  Within this approximation, one finds that $H_{g_4}$ only contributes to a shift in the electron chemical potential, which can be absorbed into a redefinition of $E_F$. Qualitative effects on the ground state can thus only come from $H_{g_1}$ and $H_{g_2}$.  In particular they can support the formation of broken-symmetry states by spontaneous admixture of the two valleys.  Because these reside in difference regions of the BZ, such admixed states represent density waves.  In our analysis we specifically examine the stability of charge density wave (CDW) and spin density wave (SDW) states.
The relevant order parameters for these have the forms
\begin{equation}\label{OP}
    \begin{split}
        & \Delta_{CDW} =(2g_1-g_2) \sum_{\mathbf q} \braket{C^\dag_{\mathbf{q} R \sigma} C_{\mathbf{q} L \sigma}} \, \textrm{ for } \sigma=\uparrow,\downarrow;
        \\&
        \Delta_{SDW}^{(+)} = -g_2 \sum_{\mathbf q} \braket{C^\dag_{\mathbf{q} R\uparrow} C_{\mathbf{q} L \downarrow} },
        \\& \Delta_{SDW}^{(-)} = -g_2 \sum_{\mathbf q} \braket{C^\dag_{\mathbf{q} R \downarrow} C_{\mathbf{q} L \uparrow} }.
    \end{split}
\end{equation} 
where $\braket{C^\dag_{\mathbf{q} \tau_1 \sigma_1} C_{\mathbf{q} \tau_2 \sigma_2}}$  represents an expectation value in the HF state.  Note that we assume the CDW order parameter $\Delta_{CDW}$ is the same for either spin, as appropriate in the absence of a Zeeman term in the Hamiltonian. For the SDW order parameter, we have introduced a helicity index $h= +(-)$ for $\tau = \sigma(-\sigma)$.

In terms of these order parameters, the resulting HF Hamiltonian can be written in the form $H_{HF}=\sum_{\mathbf k}\Psi_{\mathbf k}^\dag H_{\mathbf k} \Psi_{\mathbf k}$, where
\begin{equation}
H_{\mathbf k}= 
    \begin{pmatrix}
    \xi_{\mathbf k,R} & \Delta^*_{CDW} & 0 & \Delta^{+ *}_{SDW} \\
    \Delta_{CDW} & \xi_{\mathbf k,L} & \Delta^{-}_{SDW} & 0\\
    0 & \Delta^{- *}_{SDW}  & \xi_{\mathbf k,R} & \Delta^*_{CDW} \\
    \Delta^{+}_{SDW} & 0 & \Delta_{CDW} & \xi_{\mathbf k,L} \\
    \end{pmatrix},
\end{equation}
and
\begin{equation}
    \Psi_{\mathbf k} = \begin{pmatrix}
     C_{{\mathbf k},R,\uparrow} \\
     C_{{\mathbf k},L,\uparrow} \\
     C_{{\mathbf k},R,\downarrow} \\
     C_{{\mathbf k},L,\downarrow}
    \end{pmatrix}\, .
\end{equation}
Solutions to the HF equations involve finding values of the order parameters in which the values of $\Delta_{CDW}$ and $\Delta_{SDW}^{\pm}$ entering $H_{\mathbf k}$ are consistent with their values as ground state expectation values of $H_{HF}$, given by Eqs. \ref{OP}.  For such solutions we find that the $\Delta^{+}_{SDW}$ has the same magnitude as $\Delta^{-}_{SDW}$, but that they may have different phases.
We thus write $\Delta^{+}_{SDW} \equiv \Delta_{SDW} =  \Delta^{-}_{SDW}e^{i \phi}$. 
We then have
\begin{equation}
H_{\mathbf k} = 
    \begin{pmatrix}
    \xi_{{\mathbf k},R} & \Delta^*_{CDW} & 0 & \Delta^*_{SDW} \\
    \Delta_{CDW} & \xi_{{\mathbf k},L} & \Delta_{SDW}e^{-i\phi} & 0\\
    0 & \Delta^*_{SDW}e^{i\phi}  & \xi_{{\mathbf k},R} & \Delta^*_{CDW} \\
    \Delta_{SDW} & 0 & \Delta_{CDW} & \xi_{{\mathbf k},L} \\
    \end{pmatrix}\, .
\end{equation}

As a step towards diagonalizing this Hamiltonian, we bring it to a block diagonal form via the unitary transformation $U^\dag H_{B,\mathbf k} U =H_{\mathbf k}$, where 
\begin{equation}
        U 
        =\frac{1}{\sqrt{2}} 
        \begin{pmatrix}
            I_2 & e^{-\frac{i \phi}{2}}I_2  \\
            I_2& -e^{-\frac{i \phi}{2}}I_2
        \end{pmatrix}
\end{equation}
where $I_2$ is the 2x2 identity matrix. Introducing the definitions 
\begin{equation}\label{OP S-A}
    \Delta_{s/a} = \Delta_{CDW} \pm e^{-\frac{i\phi}{2}}\Delta_{SDW}\, ,
\end{equation}
the Hamiltonian can be recast in the form
\begin{equation}
    H_{HF}=\sum_{\mathbf k} \gamma^\dag_{\mathbf k}H_{B,\mathbf k}\gamma_{\mathbf k}, 
\end{equation}
where
\begin{equation}\label{Block-diag}
    H_{B,\mathbf k} =\begin{pmatrix}
     \xi_{{\mathbf k},R} & \Delta^*_{a} & 0 & 0 \\
     \Delta_{a} & \xi_{{\mathbf k},L} & 0 & 0 \\
     0 & 0 & \xi_{{\mathbf k},R} & \Delta^*_{s} \\
     0 & 0 & \Delta_{s} & \xi_{{\mathbf k},L} \\ 
    \end{pmatrix}
\end{equation}
and
\begin{equation}
    {\gamma}_{\mathbf k} = 
    U \Psi_{\mathbf k} = \frac{1}{\sqrt{2}}\begin{pmatrix}
     C_{{\mathbf k},R,\uparrow}+e^{-\frac{i\phi}{2}}C_{{\mathbf k},R,\downarrow} \\
     C_{{\mathbf k},L,\uparrow}+e^{-\frac{i\phi}{2}}C_{{\mathbf k},L,\uparrow} \\
     C_{{\mathbf k},R,\uparrow}-e^{-\frac{i\phi}{2}}C_{{\mathbf k},R,\downarrow} \\
     C_{{\mathbf k},L,\uparrow}-e^{-\frac{i\phi}{2}}C_{{\mathbf k},L,\downarrow}\\
    \end{pmatrix}\, . 
\end{equation}
Each block is then diagonalized independently by a Bogoliubov transformation yielding the eigenvalues
\begin{equation}\label{eigen coexist}
    \lambda_{\mathbf k,\pm}(\Delta_\nu) = \bar \xi_{\mathbf k} \pm \sqrt{\delta \xi_{\mathbf k}^2+|\Delta_{\nu}|^2}
\end{equation}
for $\nu=a,s$, in which
\begin{equation}\label{eq:xi_bar_delta_xi}
    \begin{split}
        &\bar \xi_{\mathbf k} =\frac{1}{2}(\xi_{\mathbf k,R}+\xi_{\mathbf k,L}) \, ,
        \\&\delta \xi_{\mathbf k} =\frac{1}{2}(\xi_{\mathbf k,R}-\xi_{\mathbf k,L}) \, .
    \end{split}
\end{equation}
Employing these results to evaluate the expectation values in Eq. (\ref{OP}), we obtain the self-consistency equations
\begin{equation}\label{eq:CDW and SDW}
    \begin{split}
        & \Delta_{SDW} =\frac{g_2}{2}e^{\frac{i \phi}{2}}\left\{\Delta_sI(\Delta_s)-\Delta_a I(\Delta_a)\right\},
        \\& \Delta_{CDW} = \frac{(g_2-2g_1)}{2}\left\{\Delta_sI(\Delta_s)+\Delta_a I(\Delta_a)\right\},
    \end{split}
\end{equation}
where
\begin{equation}\label{I1,I2}
    I(\Delta_\nu) \equiv \frac{1}{2}\int_{} \frac{d^2k}{(2\pi)^2}\frac{f(\lambda_{\mathbf k,-}(\Delta_\nu))-f(\lambda_{\mathbf k,+}(\Delta_\nu))}{\sqrt{(\delta \xi_{\mathbf k})^2+|\Delta_\nu|^2}}\, ;
\end{equation}
here $f(\varepsilon)$ is the Fermi-Dirac distribution, and $\mathbf{k}$ has been written in units where the volume of the real space unit cell is 1.
Note that $I(\Delta_\nu)$ depends only on the magnitude $|\Delta_\nu|$.  

Eqs. (\ref{eq:CDW and SDW}) can be written in a more symmetric form by recasting them fully in terms of $\Delta_s$ and $\Delta_a$.
Using Eq. (\ref{OP S-A}), one finds

\begin{equation}\label{eq:SCF coexistance}
    \begin{split}
        &\Delta_s = (g_2-g_1) \Delta_s I(\Delta_s)-g_1 \Delta_a I(\Delta_a),
        \\&\Delta_a = (g_2-g_1) \Delta_a I(\Delta_a)-g_1\Delta_s I(\Delta_s)\, ,
    \end{split}
\end{equation}
which can be expressed in a matrix form,
\begin{equation}\label{CDW_SDW matrix form}
\hat A\begin{pmatrix}
        \Delta_s \\ \Delta_a
    \end{pmatrix}=0,
\end{equation}
where 
\begin{equation}
\hat A=\begin{pmatrix}
    (g_{2}-g_{1})I(\Delta_{s})-1 & -g_{1}I(\Delta_{a})\\
    -g_{1}I(\Delta_{s}) & (g_{2}-g_{1})I(\Delta_{a})-1
\end{pmatrix} \; .
\end{equation}
Non-trivial solutions to Eq. (\ref{CDW_SDW matrix form}) require $\det(\hat A)=0$, leading to
\begin{equation}\label{eq:I(ds,da)}
    \begin{split}
    & \Delta_{a}=\left[I(\Delta_{s})\left(g_2^2-2 g_1 g_2\right) -\left(g_2-g_1\right)\right]\frac{\Delta_{s}}{g_1} \equiv \mathcal{F}(\Delta_{s}),
    \\& \Delta_{s}=\left[I(\Delta_{a})\left(g_2^2-2 g_1 g_2\right) -\left(g_2-g_1\right)\right]\frac{\Delta_{a}}{g_1} \equiv \mathcal{F}(\Delta_{a})\; .
    \end{split}
\end{equation}
Independent equations for $\Delta_s$ and $\Delta_a$ can be generated by noting $\Delta_i=\mathcal{F}(\mathcal{F}(\Delta_i))=\mathcal{G}(\Delta_i)$, with $i=a$ or $s$. Since both sectors have the exact same self-consistent equation, we expect the solutions to obey $\Delta_a=\pm \Delta_{s}$.
This observation indicates that either $\Delta_{CDW} = 0$ or $\Delta_{SDW} = 0$.  Thus, we expect the system to form either a charge density wave or a spin density wave, but not an admixture of them.

Of these two possibilities, we expect that $\Delta_{CDW} = 0$ [i.e., spin density wave (SDW) order] will be realized, as it is lower in energy for $g_1>0 $ (repulsive interactions) \cite{Overhauser_1960,Overhauser_1962}. 
Adopting this assumption, we conclude that the HF Hamiltonian (Eq. \ref{Block-diag}) can be recast in helicity sectors, in the form $H_{MF} = \sum_{\mathbf k, h} \tilde\Psi^\dag_{\mathbf k} \tilde H_{\mathbf k}\tilde\Psi_{\mathbf k}$, where
\begin{equation}
    \tilde H_{\mathbf k}=\begin{pmatrix}
    H_{\mathbf k, +} & 0 \\ 0 & H_{\mathbf k, -}
\end{pmatrix},
\end{equation}
\begin{equation}\label{H_k}
     H_{\mathbf k,h}=\begin{pmatrix}
    \xi_{\mathbf k,R} & \Delta_h \\
    \Delta_h^* & \xi_{\mathbf k,L}
\end{pmatrix},
\end{equation}
and
\begin{equation}
\tilde\Psi_{\mathbf k}^\dag=\begin{pmatrix}
C_{\mathbf k, R, \uparrow}^\dag & C_{\mathbf k, L, \downarrow}^\dag & C_{\mathbf k, R, \downarrow}^\dag & C_{\mathbf k, L, \uparrow}^\dag 
\end{pmatrix}.
\end{equation} 
Using Eqs. (\ref{OP S-A}) and (\ref{eq:SCF coexistance}) for the case
$\Delta_{CDW}=0$, 
we arrive at a self-consistent gap equation for an order parameter with definite helicity $h$, in the form
\begin{equation}\label{gap_eq_Delta_h}
    \Delta_{h}=g_2\Delta_{h}I(\Delta_h).
\end{equation}
Although this equation allows for one $\Delta_h$ to vanish while the other does not, we find that the lowest energy solutions always occur for $|\Delta_+|=|\Delta_-|\equiv\Delta$.  Note that the relative phase of the two order parameters may take any value, without affecting the energy of the state.

While Eq. (\ref{gap_eq_Delta_h}) has the apparent form of a standard gap equation, it should be emphasized that its solution depends self-consistently on one more variational parameter in addition to $\Delta$: the nesting vector $\mathbf{Q}$ as parameterized by $Q_y$. This dependence is encoded in the single-body energies Eq. (\ref{eq:xi_def}) which determine the integral $I(\Delta)$ (see Eqs. (\ref{eq:xi_bar_delta_xi}), (\ref{I1,I2})). In the next section we present the results of this variational calculation.   

\begin{figure}
    \centering
    \includegraphics[width=1\linewidth]{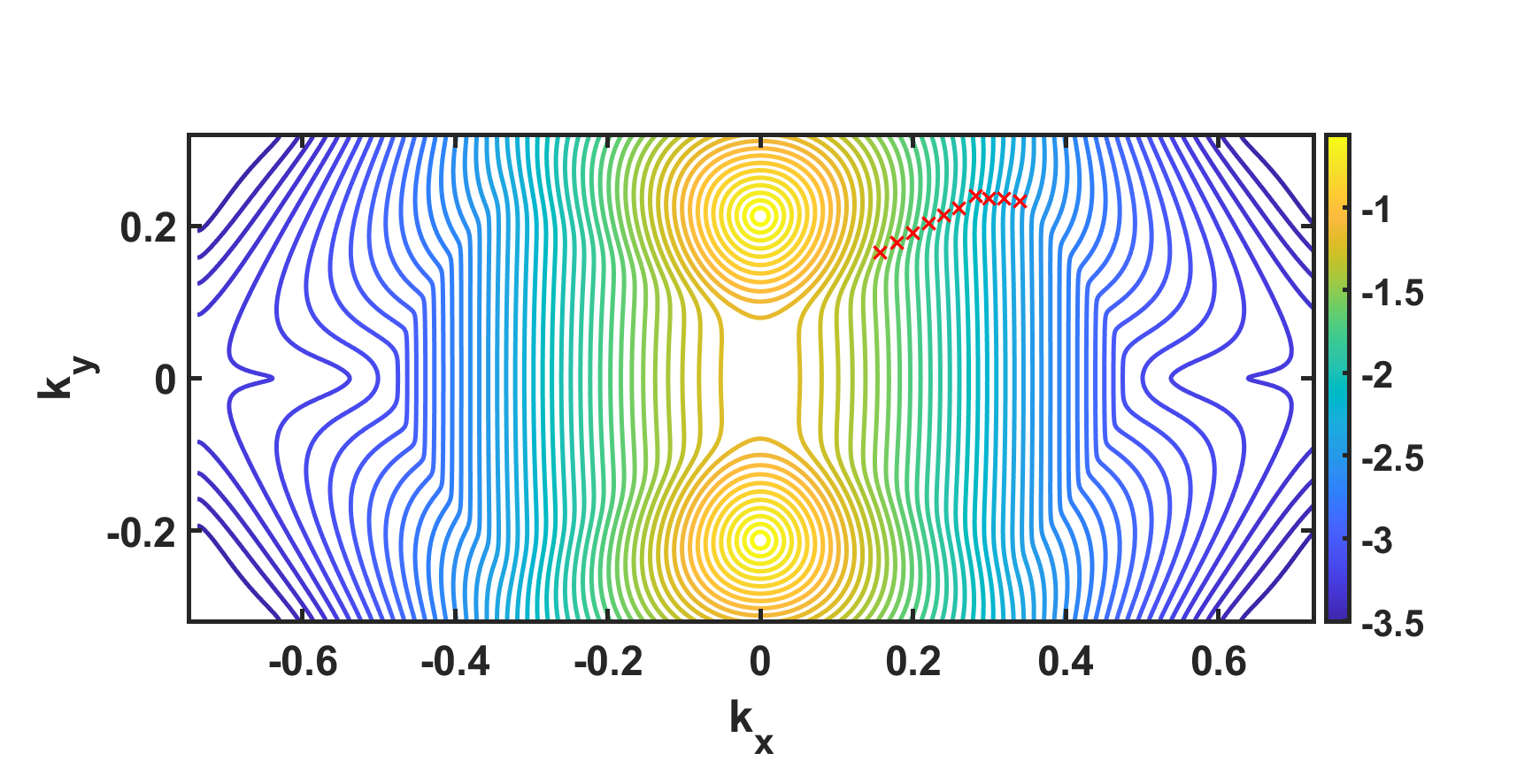}
    \caption{Numerically evaluated nesting vector.  For each $E_F$ (marked by the color scale in $\text{eV}$), the red cross marks $\frac{1}{2}\mathbf{Q}$ on the Fermi surface (the one of four hot-spots corresponding to the quarter-plane $Q_x,Q_y>0$). }
    \label{fig:nesting_vec}
\end{figure}

\begin{figure}
    \includegraphics[width=0.45\textwidth]{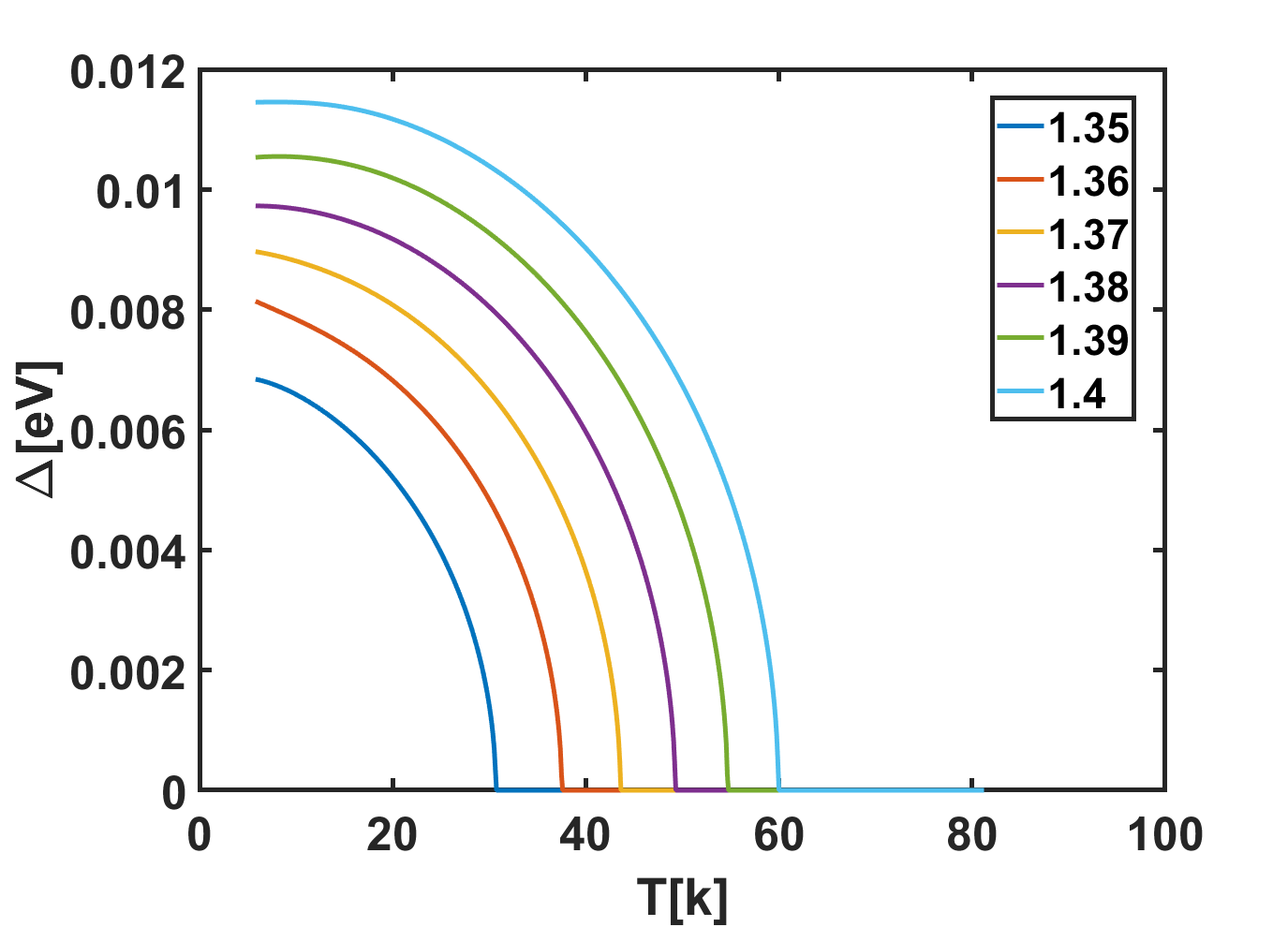}
    \caption{A numerical calculation of the gap $\Delta(T)$ for various values of the interaction strength $G$ (in units of $\text{eV}(\text{\AA})^2$) presented in the legend. Here $E_F=-2.2\,\text{eV}$, $\frac{1}{2}\mathbf{Q} =0.3 \hat{x} +0.235\hat{y}$.}
    \label{gap-T}
\end{figure}

\begin{figure}
\centering
\begin{subfigure}{0.45\textwidth}
\caption{}
    \includegraphics[width=1\textwidth]{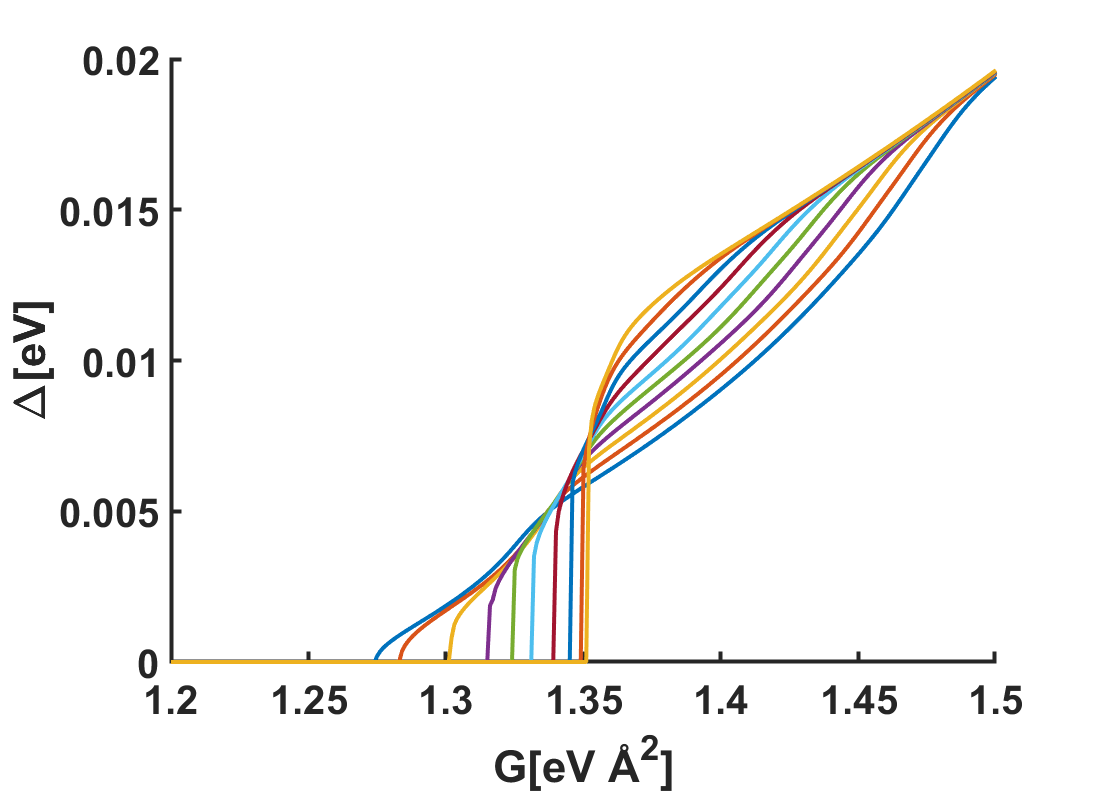}
    \label{D-G}
\end{subfigure}
\begin{subfigure}{0.45\textwidth}
\caption{}
    \includegraphics[width=1\textwidth]{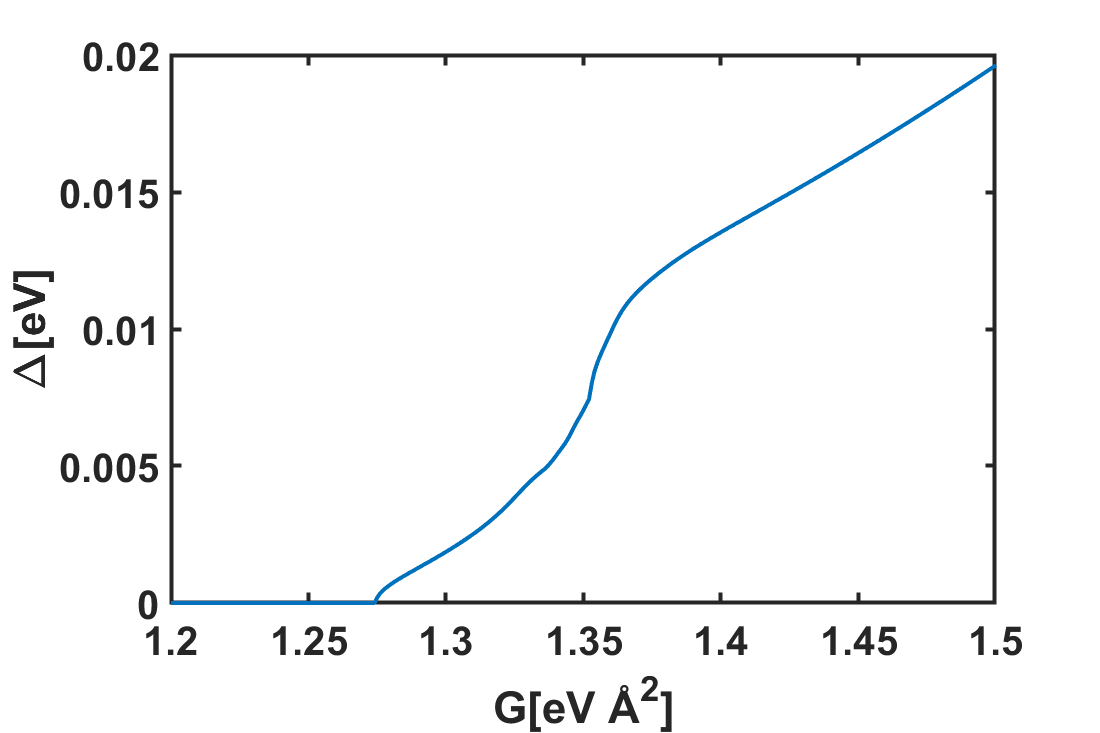}
    \label{}
\end{subfigure}
    \caption{The gap $\Delta(G)$ at $T\to0$ and $E_F=-2.2\,\text{eV}$. (a) Solution of the gap equation calculated for 10 evenly spaced choices of $\frac{1}{2}Q_y$ close to the optimal value $\frac12 Q_y=0.235$ in the interval $[0.2323,0.2387]$.
    (b) The maximal value of $\Delta$ obtained from the previous plot as a function of $G$. From this, we observe that there is a critical value $G_c$ above which there is a finite solution to the gap equation.}
    \label{fig: D_G multiple Q_y}
\end{figure}

\section{Principal Results and Phase Diagram}
\label{Sec:results}

The mean-field analysis described above suggests that, provided Eq. (\ref{gap_eq_Delta_h}) has a non-trivial solution, the system develops a SDW instability which, owing to the characteristic nesting vector $\mathbf Q$, exhibits a striped pattern. We now explore its behavior upon tuning of the model parameters, which besides the temperature $T$ include the interaction strength $G\equiv g_2/(2\pi)^2$ and the Fermi energy $E_F$; the latter is in principle tunable by a gate voltage, thus serving as a convenient control parameter in an experimental realization. To this end, we adopt the single-body energy spectrum obtained from the tight-binding model (Sec. \ref{ssec: T.B.}) as input to Eqs. (\ref{eq:xi_bar_delta_xi}), (\ref{I1,I2}), and numerically solve the gap equation (\ref{gap_eq_Delta_h}) for different values of $G$ and $E_F$. This enables us to evaluate $\Delta$ at arbitrary $T$, and particularly to construct a phase diagram of the ground-state ($T=0$) in the $E_F-G$ plane.

The primary challenge in the calculation involves the identification of the $\hat y$-component of the optimal nesting vector $\mathbf Q$. As mentioned in the previous subsection, we consider $Q_y$ as a variational parameter and find its optimal value for each set of $G$ and $E_F$ by 
minimizing the ground state energy (see typical traces of $\Delta$ vs. $Q_y$ in Appendix \ref{AppA}). Note that in the gap equation (\ref{gap_eq_Delta_h}), $Q_y$ also dictates the momentum cut-off on the integral Eq. (\ref{I1,I2}), as it defines the parallelogram-shaped reconstructed BZ (Fig. \ref{rec_BZ}).

Numerically, we find that the value of the optimal $Q_y$ is most strongly dependent on $E_F$. Indeed, this reflects sensitivity to the detailed structure of a particular Fermi surface. More concretely, we find that the Fermi points $\frac{1}{2} \mathbf Q$ approximately coincide with inflection points in the Fermi curves, as clearly indicated in Fig. \ref{fig:nesting_vec}.  
Notably, by symmetry of the band-structure, the resulting optimal $\mathbf Q$ depends only on the magnitude of $Q_y$ for a given $Q_x$. As a result, there are two degenerate but inequivalent solutions to the emergent nesting vector, $\mathbf{Q}_\pm=|Q_x|\hat{x}\pm |Q_y|\hat{y}$, corresponding to distinct orientations of the nematic order expected in the observable striped SDW pattern related by mirror symmetry across the $x$-axis. As such, the striped SDW state breaks a $Z_2$ symmetry of the lattice structure in addition to the broken continuous symmetries in the spin and valley sectors ($SU(2)$ and $U(1)$ respectively).     

Fig. \ref{gap-T} illustrates typical solutions for the gap as function of  $T$ across a range of interaction parameters $G$, for one particular choice of $E_F$. It presents a characteristic mean-field like suppression of the gap at a $G$-dependent critical temperature $T_c$. Interestingly, however, a threshold value of $G$ is needed to obtain a non-zero order parameter even at $T=0$. This is illustrated in Fig. \ref{D-G}. 

The existence of such a threshold can be explained with the aid of a simple model for the single-body energies $\xi_{\mathbf k,\tau}$.
We consider an approximate dispersion law $\xi_{\mathbf k,\tau}$ vs. $\mathbf k$ in the vicinity of the Fermi points $\frac{1}{2} \tau\mathbf Q$ which assumes a form similar to Eq. (\ref{toy_dispersion}); however, accounting for the curvature of the  Fermi surfaces and their displacement from the $\Gamma$ point, we define a local coordinate system $(k_x,k_y)$ for the deviation from $\frac{1}{2} \tau\mathbf Q$ where $k_x$ is perpendicular to the curve and $k_y$ tangential. With respect to these axes, the dispersion is assumed to be  
\begin{equation}
     \xi_{\mathbf k,\tau}\approx -\tau v k_x+ak_y^\alpha
     \label{xi_k_toy}
\end{equation} 
where $\alpha>1$. Substituting in Eqs. (\ref{eigen coexist}), (\ref{eq:xi_bar_delta_xi}),  
the energy eigenvalues become
\begin{equation}
    \lambda_{\mathbf k, \pm} = a k_y^\alpha\pm \sqrt{v^2 k_x^2+\Delta^2}\, .
    \label{lambda_k_toy}
\end{equation}

We next consider the $T\to 0$ limit of the self-consistent equation (\ref{gap_eq_Delta_h}):
\begin{equation}
    1 = \frac{G}{2}\int_{BZ}d^2k \frac{[\Theta(-\lambda_{\mathbf k, -}) -\Theta(-\lambda_{\mathbf k,+}) ]}{\sqrt{\delta\xi^2_{\mathbf k}+\Delta^2}},
\label{gap_eq_T0}
\end{equation}
where $\Theta(x)$ is the Heaviside step function. Defining
\begin{equation}
    F(\Delta)\equiv
    \frac{1}{2}\int_{BZ}d^2k \frac{[\Theta(-\lambda_{\mathbf k, -}) -\Theta(-\lambda_{\mathbf k,+}) ]}{\sqrt{\delta\xi^2_{\mathbf k}+\Delta^2}}\, ,
    \label{F_Delta_def}
\end{equation}
Eq. (\ref{gap_eq_T0}) reduces to 
\begin{equation}
    F(\Delta)=\frac{1}{G}\, .
    \label{F_Delta_G}
\end{equation}
The toy model dispersions Eqs. (\ref{xi_k_toy}), (\ref{lambda_k_toy}) now enable us to express the function $F(\Delta)$ as a straightforward integral
\begin{equation}
     F(\Delta) = \frac{2}{a^{\frac{1}{\alpha}}}
     \int_{0}^{k_c}d k_x (v^2k_x^2+\Delta^2)^{\frac{1-\alpha}{2\alpha}}  
\end{equation}
where $k_c$ is an upper cut-off on $k_x$ set by the zone boundary. We note that for arbitrary $\alpha>1$, $F(\Delta)$ is a monotonically decreasing function
bounded by its finite value at $\Delta=0$. Eq. (\ref{F_Delta_G}) then implies that there is a critical value of $G$, $G_c=\frac{1}{F(0)}$, below which it cannot be satisfied. As a result, a non-trivial solution $\Delta\neq 0$ is manifest only for $G > G_c$. 

\begin{figure}[h]
    \begin{subfigure}{0.45\textwidth}
    \caption{}
        \includegraphics[width=0.9\textwidth]{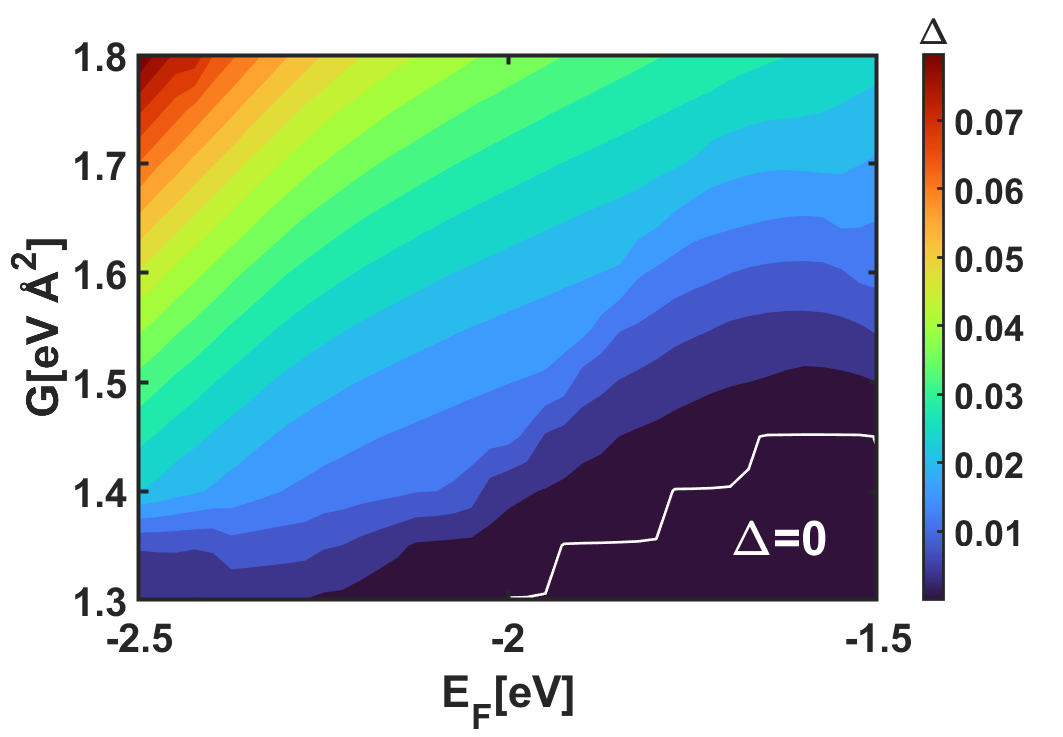}
        \label{gap graph}
    \end{subfigure}
    \begin{subfigure}{0.45\textwidth}
    \caption{}
        \includegraphics[width=0.9\textwidth]{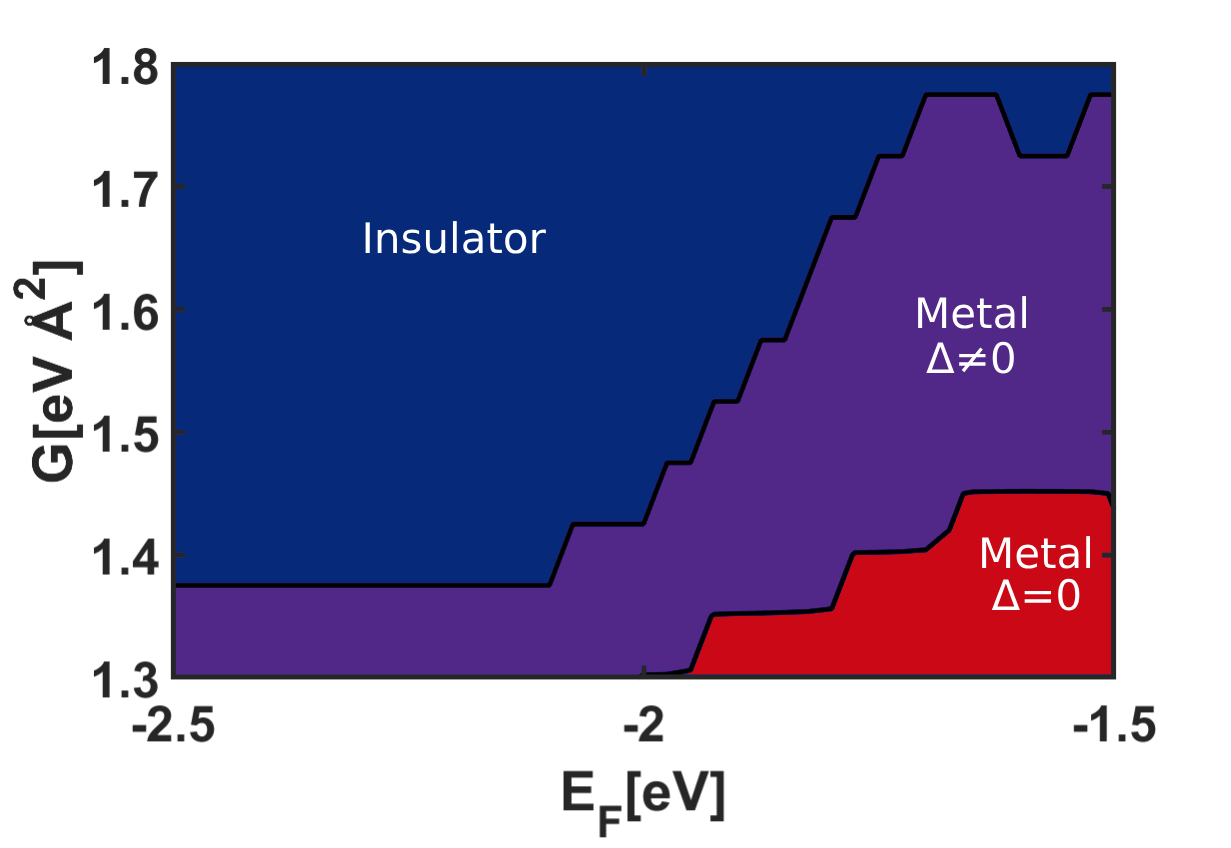}
        \label{MIT}
    \end{subfigure}
    \caption{(a) Phase diagram presenting the gap $\Delta$ as a function of the interaction strength and the Fermi energy. The transition line to the SDW phase is marked in white. (b) The two-stage metal-to-insulator transition; red represent the metallic disordered phase, purple is the metallic SDW phase, and blue is the SDW insulating phase.}
    \label{fig: Phase Diagrams}
\end{figure}

Repeating our numerical analysis of $\Delta$ vs. $G$ for a wide range of $E_F$ yields the $T=0$ phase diagram Fig. \ref{gap graph}. It shows that the SDW phase is stable for sufficiently large interaction parameters $G$, with critical values $G_c$ which depend on $E_F$, with the latter in the range $-2.5\,\text{eV}\leq E_F\leq -1.5\,\text{eV}$.
This corresponds to the regime where the two parts of the Fermi surface can be nearly overlaid by a nesting vector. 

\begin{figure}
    \begin{subfigure}{0.45\textwidth}
    \caption{}
        \includegraphics[width=0.9\textwidth]{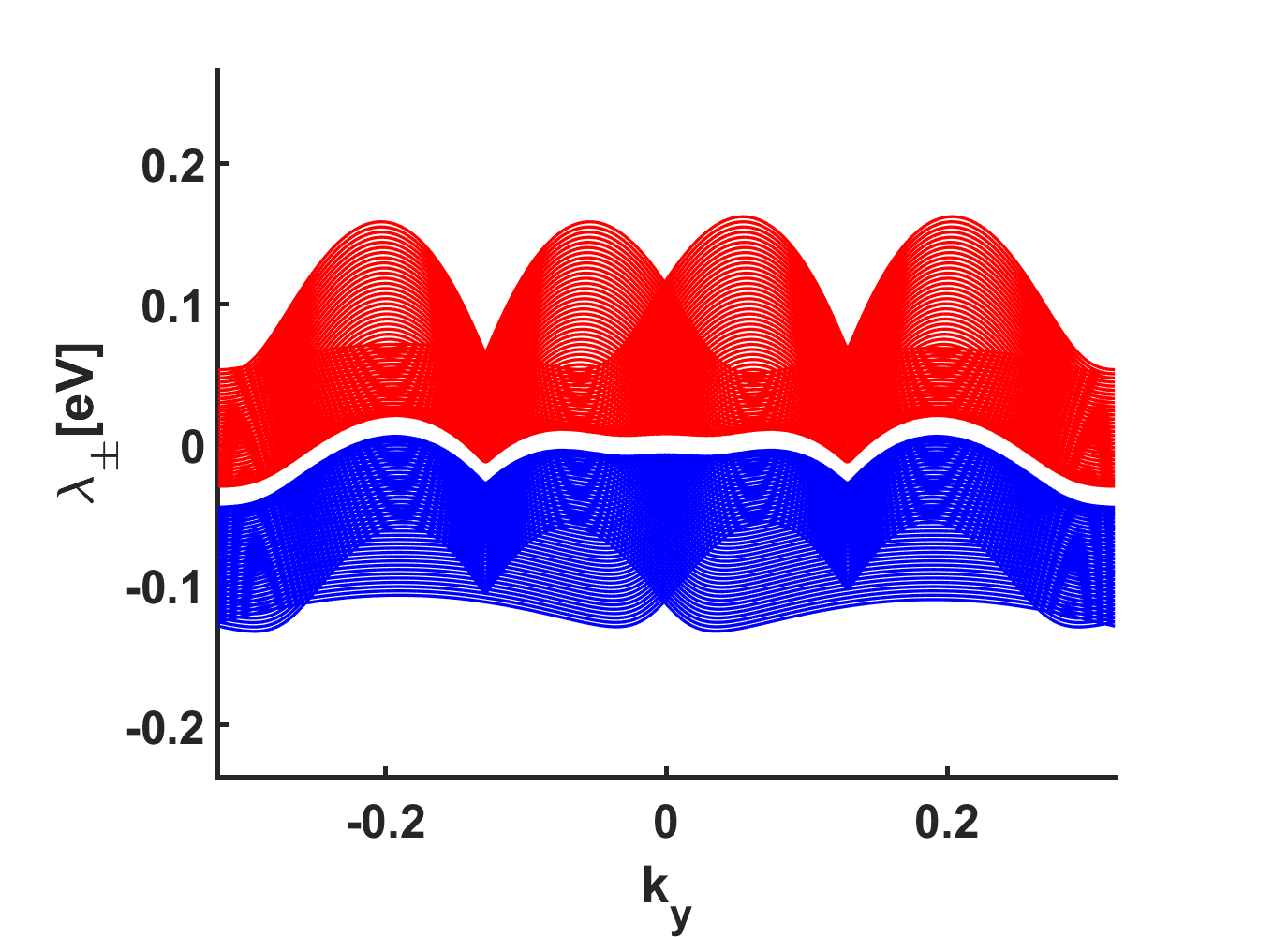}
        \label{lambda_M}
    \end{subfigure}
    \begin{subfigure}{0.45\textwidth}
    \caption{}
        \includegraphics[width=0.9\textwidth]{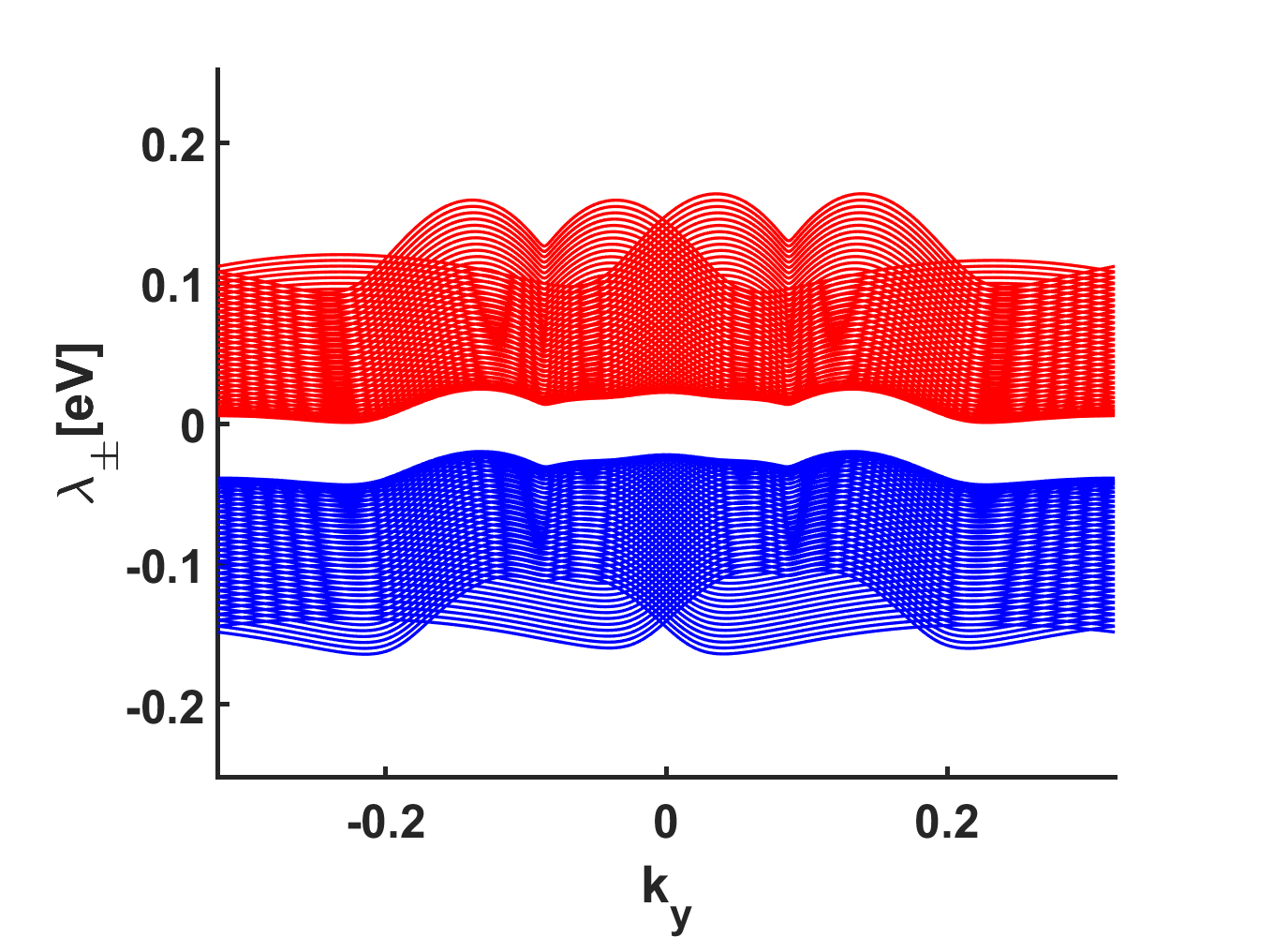}
        \label{lambda_I}
    \end{subfigure}
    \caption{Projection of the excitation energy spectra $\lambda_{\mathbf{k},+}$ (red) and $\lambda_{\mathbf{k},-}$ (blue) vs. $\mathbf{k}$ for (a) $E_F=-1.7\, \text{eV}$, $G=1.55\, \text{eV}(\text{\AA})^2$ and nesting vector $\frac12\mathbf{Q}=(0.1998,0.1903)$ (the metallic SDW phase) and for (b) $E_F=-2\, \text{eV}$, $G=1.6\, \text{eV}(\text{\AA})^2$ and nesting vector $\frac12\mathbf{Q}=(0.2620,0.2322)$ (the insulating SDW phase). }
    \label{fig:Lambda_pm}
\end{figure}

Further inspection of the excitation spectrum (Eq. \ref{eigen coexist}) in the ordered regime where $\Delta\not =0$ shows that it in fact supports two distinct phases: for a given $E_F$, the Fermi surface fully disappears only above a finite threshold value of $\Delta$. Below this threshold, the system    
features an intermediate metallic phase where the quasi-particle energy bands $\lambda_{\mathbf{k},\pm}(\Delta)$ overlap, meaning that $\max(\lambda_-)>\min(\lambda_+)$.  This is illustrated in Fig. \ref{lambda_M}, and it ensures that $E_F$ will lie inside at least one band. This metallic region is highlighted in purple as shown in Fig. \ref{MIT}. Above a critical line in the $E_F-G$ plane, a metal-insulator transition occurs to an insulting phase (blue region in Fig. \ref{MIT}) where the Fermi surface is fully gapped (see Fig. \ref{lambda_I}).
In both phases, the non-vanishing order parameter $\Delta$ suggests a broken symmetry in the  ground state, most naturally described as a unidirectional SDW.

\begin{figure}
    \centering
    \includegraphics[width=1\linewidth]{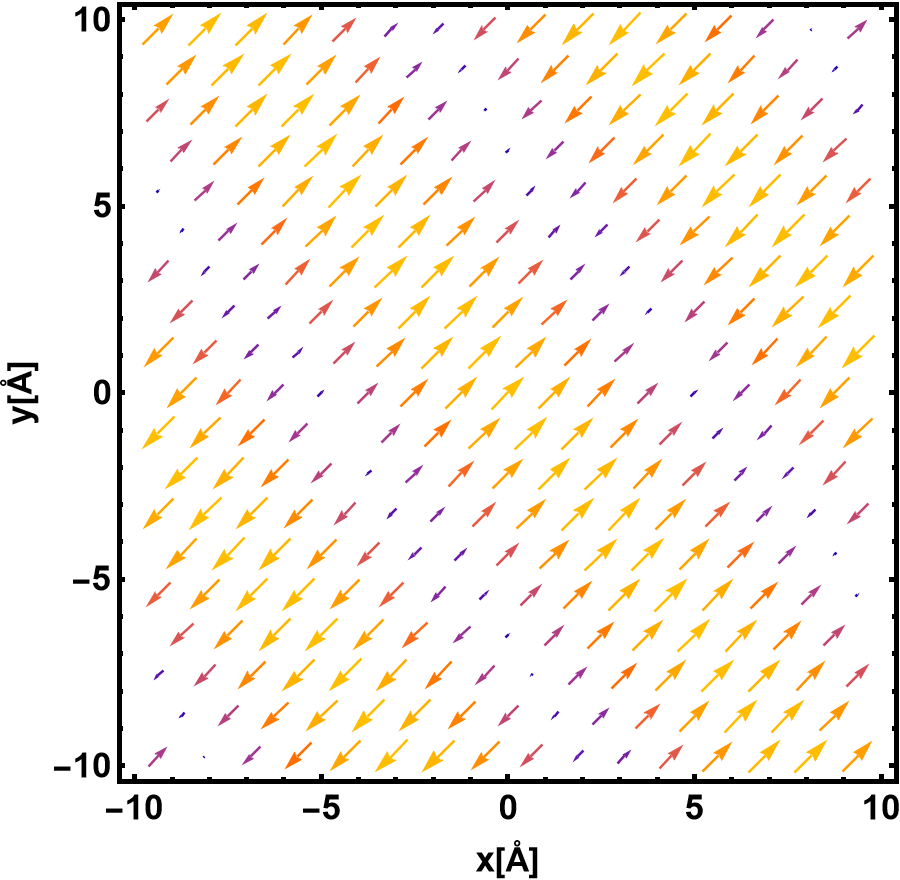}
    \caption{Representation of the SDW phase in real space for $\frac{1}{2}Q=(0.3,0.235)$ and the specific choice of phases $\alpha=-\beta=\frac{\pi}{4}$ (see Appendix \ref{spin_exVal} for details).}
    \label{fig:realSpaceSDW}
\end{figure}

To illustrate the manifestation of the SDW order in real-space, we compute expectation values of the spin operator components in the ground state for a given value of the complex order parameters $\Delta^{(\pm)}$ (see Appendix \ref{spin_exVal}). This yields
\begin{equation}
\braket{S^+}=\braket{S_x}+i\braket{S_y}\sim e^{i \varphi}\cos(\mathbf Q \cdot \mathbf r+\phi)
\label{S+ExpVal_mt}
\end{equation}
where $\varphi$ and $\phi$ are, respectively, the relative and average phases of $\Delta^{(\pm)}$.
The resulting pattern is depicted in Fig. \ref{fig:realSpaceSDW}. 
It is worth noting that this specific spin pattern is the result of two spiral-SDW components ($\Delta_h$ with $h=\pm1$)  counter-rotating in the $x-y$ plane, characterize by Eq. (\ref{Eq:SSDW}). This configuration forms ferromagnetic stripes oriented  perpendicular to $\mathbf{Q}$, with the spin directions of these stripes flipping as one moves along the $\mathbf{Q}$ direction. Note that within the $S_x-S_y$ plane, the spins in a stripe point in an arbitrary direction dictated by the phase difference of the two modes. Finally, we point out that our choice of the spin $z$-direction is arbitrary, and that $\braket{S_z}$ vanishes in this state.  However, because our Hamiltonian is fully $SU(2)$ symmetric in the spin variables, any global rotation of all the spins will yield a different spin-stripe state at the same energy, in which 
$\langle S_z \rangle$  need not vanish.
A more detailed discussion of how we arrive at the spin expectation values illustrated in Fig. \ref{fig:realSpaceSDW} is provided in Appendix \ref{spin_exVal}.

As a final comment, we note that 
the SDW phase obtained above exhibit a spontaneous breaking of time-reversal symmetry. In closing our mean-field analysis, we consider the effect of explicit time-reversal symmetry-breaking in the form of an external magnetic field $B$. This couples to the electron spins via a Zeeman term added to the Hamiltonian,
\begin{equation}
    H_z = -B \hat s_z = -B(\hat n_{\uparrow}-\hat n_{\downarrow}),
\end{equation}
where $B$ is the magnetic field, $\hat{n}_{\uparrow \, (\downarrow)}$ is the number operator for up-  (down-) spin electrons, and
we have adopted units where $g\mu_B$ (with $g$ the Land\'e $g$-factor, and $\mu_B$ the Bohr magneton) is 1. This induces a spin dependence in the single body energies: $\xi_{\mathbf k,\tau} \to \xi_{\mathbf k, \tau, \sigma}=\xi_{\mathbf k,\tau}-B\sigma$, thus modifying the MF Hamiltonian [Eq. (\ref{H_k})] yielding the helicity blocks 
\begin{equation}
    H_{\mathbf k,h} = \begin{pmatrix}
     \xi_{\mathbf k,R}-B h && \Delta_{h}\\
     \Delta^{*}_{h} && \xi_{\mathbf k,L}+B h
    \end{pmatrix}.
\end{equation}
The resulting self-consistent gap equations now take the form
\begin{equation}\label{gap_eq_B}
        \Delta_{h} = \frac{g_2}{2}\int\frac{d^2k}{(2\pi)^2}\frac{\Delta_{h}[f(\lambda_{\mathbf k,-})-f(\lambda_{\mathbf k,+})]}{\sqrt{|\Delta_{h}|^2+(\delta\xi_{\mathbf k}-B h)^2}}
         \, .
\end{equation}
Note that the gap openings are now at spin-dependent Fermi surfaces ($\mathbf k \ne 0$) satisfying $\delta\xi_{\mathbf k}=Bh$. For physically accessible values of $|B|$, this occurs at a small shift in momentum $ \mathbf k = q \hat x$ where $\xi_{\mathbf k,\tau}$ can be approximately linearized to yield $\xi_{\mathbf k,\tau} \approx -v\tau q$, and $q=\mp \frac{B}{v} $ for $h=\pm $ respectively. Our numerical solution of the gap equation (\ref{gap_eq_B}) indicates that as long as the shift $q$ leaves the Fermi surfaces within the quasi-1D regime of the BZ, the solution manifests a negligible change compared to the $B=0$ case.

\section{Discussion and outlook}\label{Discussion}
In this paper we have presented a study of a graphene-black phosphorus (G/BP) heterostructure.  The electronic band-structure of this system supports nearly straight constant energy contours over a range of Fermi energies, which are excellent candidates for nesting instabilities. This behavior, which stems from a strong hybridization of the top valence band of BP with the lower Dirac cones of the graphene layer, is captured by a simple tight-binding model for the single-body Hamiltonian. Introducing a minimal model for the interactions, we perform a mean-field (MF) analysis of possible broken-symmetry ground states.  
We found that instabilities indeed set in for strong enough interactions, and that the lowest energy configurations involve unidirectional striped spin density wave (SDW) states.

An interesting aspect of the emergent SDW order is that its characteristic wave-vector $\mathbf{Q}$ is dependent on the parameters of the system and most prominently on the Fermi energy $E_F$. It is typically incommensurate with the lattice, and encodes inflection-point features in the Fermi surface that are therefore sensitive also to details of the single-body Hamiltonian (e.g., the tight-binding parameters $\epsilon_G$ and inter-layer hopping coefficients $t_i$). As such, there are multitude ways to control the wavelength and orientation of the stripes by gating (particularly in a dual-gating setup) and strain.

While the MF analysis indicates  that the SDW instability is accompanied by the opening of a direct gap $|\Delta |$ in the electronic spectrum, a detailed study of the excitation energy bands suggests that the Fermi surface does not fully disappear for sufficiently small $|\Delta |$. Hence, the system hosts two distinct striped SDW phases, a metallic and an insulating one. We therefore anticipate a two-stage transition into the fully ordered phase, as manifested by the critical lines in the $E_F-G$ phase diagram Fig. \ref{MIT}.

The striped SDW order spontaneously breaks several symmetries of the system. Two of them are continuous symmetries: $SU(2)$ in the spin sector and $U(1)$ in the valley sector. The former broken-symmetry is encoded by the phase $\varphi$ corresponding to rotation in the $x-y$ plane of the Bloch sphere (see Eq. (\ref{S+ExpVal_mt})) and a polar angle $\vartheta$ associated with its orientation with respect to a fixed axis in space; the latter corresponds to the phase-shift $\phi$ of the stripe pattern Eq. (\ref{S+ExpVal_mt}). Additionally, a broken discrete $Z_2$ symmetry associated with a mirror symmetry relating the two degenerate choices of nesting vectors $\mathbf{Q_\pm}$ is encoded by the orientation of the stripes in the real-space $x-y$ plane. 

Preliminary studies of the fluctuations in the order parameter beyond MF-theory suggest a number of gapless collective modes characterizing the broken-symmetry phases, which exhibit highly anisotropic dispersion. Specifically, these include the spin-waves associated with fluctuations in the angles $\varphi$ and $\vartheta$, corresponding to the relative phase and amplitude mode of $\Delta^{(\pm)}$, respectively; fluctuations in the phase-shift $\phi$ correspond to a phonon mode of the quasi-1D stripes pattern. Detailed study of the dispersion laws and physical manifestations of these modes are deferred to a separate publication \cite{author2024}.

We finally comment on the prospects for experimental probing of our proposed SDW instability in a G/BP heterostructure. As mentioned in Sec. \ref{sec:introduction}, such hybrid bilayers have been fabricated and experimentally studied, e.g. in the work of Ref. \cite{liu2018tailoring}. The regime of parameters compatible with our predictions can likely be realized in a dual-gated setup, which allows an independent control of the hole-doping in the heterostructure and of the band-structure via the interlayer displacement field. We expect that by tuning the gate voltages, it is possible to drive a two-stage transition into the SDW phases marked by the emergence of anisotropic electric and thermal transport; while the former is dominated by the band-reconstruction of the conduction electrons, the latter would be a signature of the various charge-neutral collective modes, which could moreover be manipulated by magnetic fields. It is also suggestive that coupling of the charge and spin sectors (e.g. due to higher-order interaction terms allowed by the symmetries) will generate a striped pattern in the charge density as well. These could be detectable by STM measurements as reported in \cite{liu2018tailoring}. 

Notably, to access the regime of parameters which we predict to host the striped SDW phases a significant hole-doping is required. This is in principle accessible in a setup which involves ionic liquid gel dielectric \cite{wang2021recent, yu2015gate}. We anticipate that the tuning of doping level in a sufficiently wide-range would provide access to transitions from the striped SDW, favored by quasi-1D Fermi surfaces, to other types of broken-symmetry states dominated by the topologically distinct structure of the Fermi surface at lighter hole-doping of the valence band (see Fig. \ref{tight binding: b}).
This phenomenon will be explored elsewhere.      

\acknowledgments
We thank Jonathan Ruhman, Yoni Messica, Udit Khanna and Amir Dalal for fruitful discussions. We acknowledge financial support by the US-Israel Binational Science Foundation through award No. 2016130. 
DH and ES acknowledge the support of the Israel Science Foundation (ISF) Grant No. 993/19 and the US-Israel Binational Science Foundation through award No. 2018726. 
HAF acknowledges the support of the NSF through Grant No. DMR-1914451.
ES and HAF thank the Aspen Center for Physics (NSF Grant No. 1066293) for its hospitality. 

\appendix 

\begin{widetext}
\section{Numerical Analysis}
\label{AppA}
In this Appendix, we provide an overview of the numerical methods and tools used to obtain our results.
The basis of our numerical analysis is the tight-binding model, which results in the band-structure seen in Fig. \ref{tight binding}. The momentum integrals in our self-consistent equations (see Section \ref{MFA}) were approximated as discrete sums. To ensure smoother plots, all calculations were performed using finite temperature of $\sim 5 K$. The grid size in $k$-space was set to 600 by 600 points, which appeared to yield convergence in the numerical analysis. For the numerical analysis of the self consistent equation we used MATLAB's fsolve function, specifically employing the Levenberg-Marquardt algorithm.
The self-consistent equation Eq. (\ref{gap_eq_Delta_h}) was solved for selected values of $E_f$ and $G$, incorporating a momentum cutoff along the $k_x$-direction set by the reduced BZ. 

For a given value of $G$ and $E_f$, we identified the appropriate nesting vector $\mathbf{Q}$ that minimized the quasi-particle ground state energy. For a given value of $E_f$, $Q_y$ served as a variation parameter while $Q_x$ was adjusted to remain on the Fermi surface.
An interesting observation was made within a specific region of range of $G$ and $E_f$: when plotting $\Delta$ as function of $\frac{1}{2}Q_y$, we obtain a non-vanishing gap in two distinct regions of $Q_y$.  Generally this gap is almost always larger for the higher value of $\frac12 Q_y$, as illustrated in Fig. \ref{DeltaVsQy}. The values of $\Delta$ as a function of $\frac12 Q_y$ behave unsystematically below $G_c$, but not really distinguishable from zero, it is then becomes continuous above $G_c$. 

\begin{figure}[h]
    \centering
    \begin{subfigure}{1\textwidth}
    \caption{}
        \includegraphics[width=0.5\linewidth]{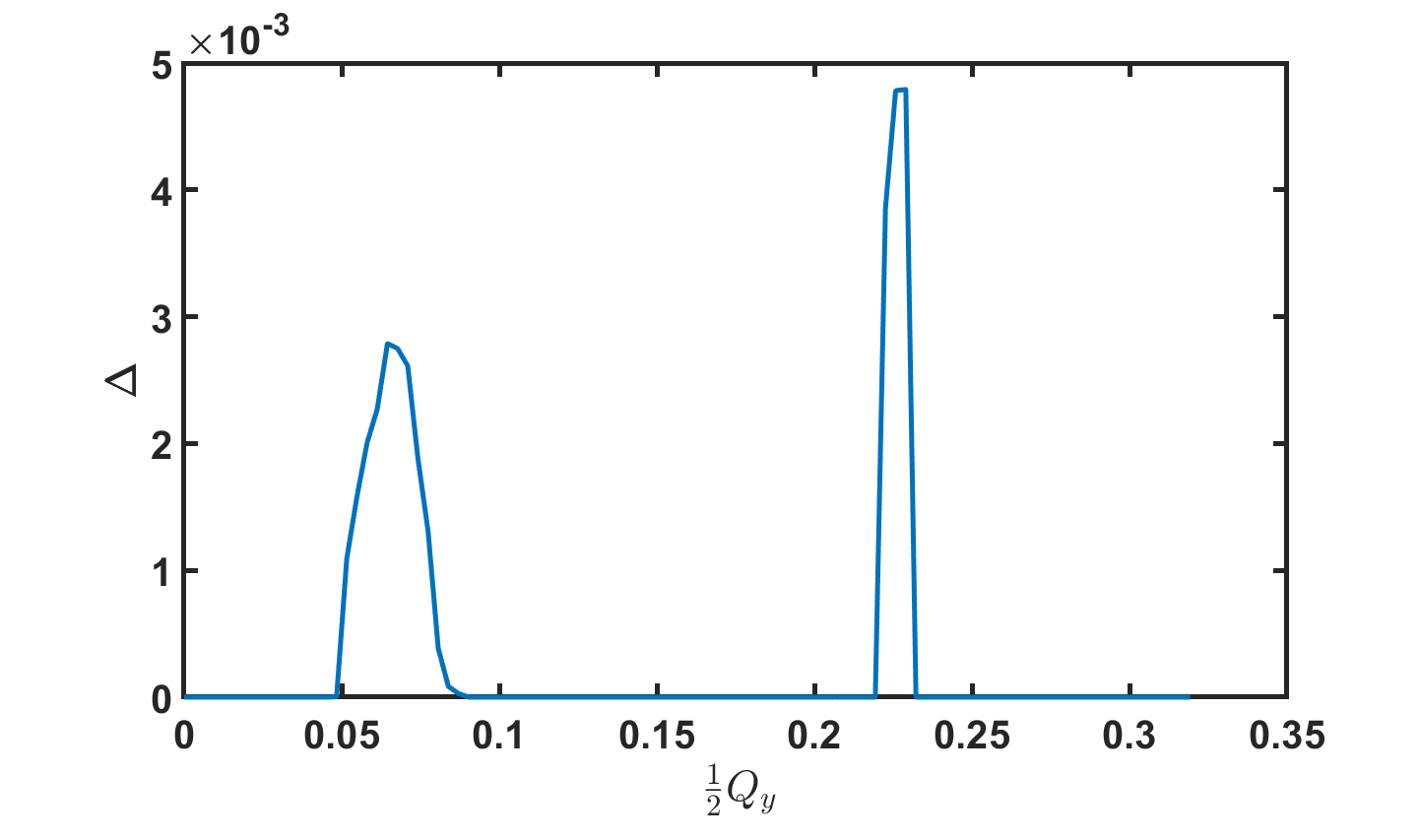}
        \label{}
    \end{subfigure}
    \begin{subfigure}{1\textwidth}
    \caption{}
        \includegraphics[width=0.5\linewidth]{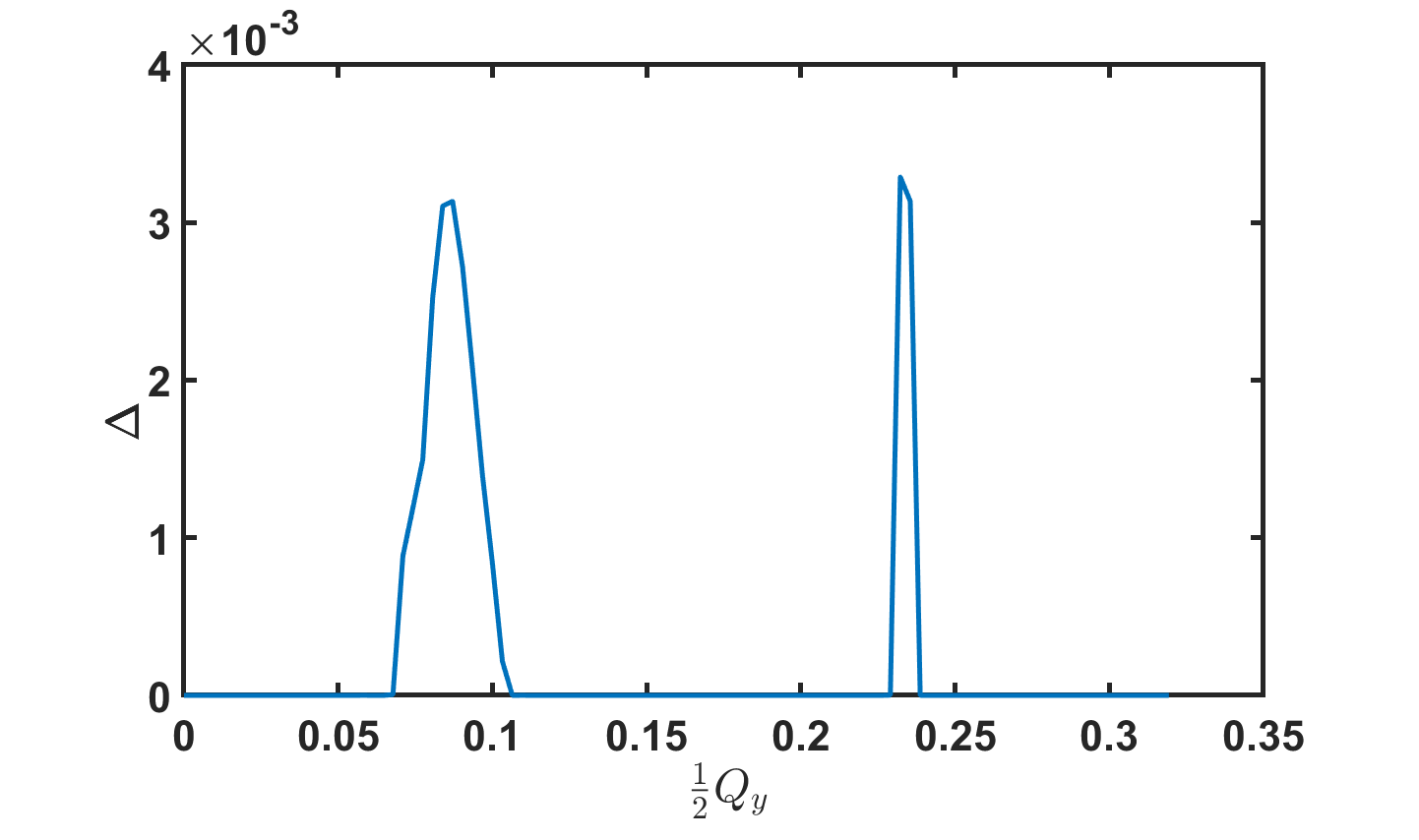}
        \label{}
    \end{subfigure}
    \caption{Numerically calculated $\Delta$ (in $\text{eV}$) as a function of $\frac{1}{2}Q_y$ (in $\text{\AA}^{-1}$) for (a) $E_F=-2\, \text{eV}$, $G=1.4\, \text{eV} (\text{\AA})^2$. (b) for $E_F=-2.25\, \text{eV}$, $G=1.3\, \text{eV} (\text{\AA})^2$. The maximal value of $\Delta$ gives the lowest ground state energy.}
    \label{DeltaVsQy}
\end{figure}

\section{Spin operator expectation values} \label{spin_exVal}
In this Appendix we derive the real-space representation of the spin density pattern supported by the striped SDW phases. To this end, we first recall the single-particle Fermion operator
\begin{equation}
    \psi_{\sigma}=\frac{1}{\sqrt{\Omega}}\sum_{\mathbf{k}}c_{\mathbf{k}.\sigma}e^{-i\mathbf{k}\mathbf{r}}\simeq \frac{1}{\sqrt{\Omega}}\sum_{\mathbf{k}\in G_{R}}c_{\mathbf{k}.\sigma}e^{-i\mathbf{k}\mathbf{r}}+\frac{1}{\sqrt{\Omega}}\sum_{\mathbf{k}\in G_{L}}c_{\mathbf{k}.\sigma}e^{-i\mathbf{k}\mathbf{r}}=\psi_{R\sigma}\left(r\right)+\psi_{L\sigma}\left(r\right),
\end{equation}
where $\Omega$ is the unit cell volume and $G_\tau$ represents the set of momenta belonging to the respective Right/Left valley.
In terms of our valley index notation,
\begin{equation}
    \label{psi2C}
    \psi_{\tau\sigma}\left(r\right)=\frac{1}{\sqrt{\Omega}}\sum_{\mathbf{k}\in G_{\tau}}c_{\mathbf{k}.\sigma}e^{-i\mathbf{k}\mathbf{r}}=\frac{1}{\sqrt{\Omega}}\sum_{\mathbf{p}}C_{\mathbf{p},\tau.\sigma}e^{-i\left(\mathbf{p}+\frac{1}{2}\tau\mathbf{Q}\right)\mathbf{r}}=\frac{e^{-i\frac{1}{2}\tau\mathbf{Q}\mathbf{r}}}{\sqrt{\Omega}}\sum_{\mathbf{p}}C_{\mathbf{p},\tau.\sigma}e^{-i\mathbf{p}\mathbf{r}}\,.
\end{equation}
The expectation value of the local $S^+$ operator (per unit cell) is given by
\begin{equation}
    \braket{S^+} =\Omega\left( \sum_{\tau=R,L}\braket{\psi^\dag_{\tau\uparrow}\psi_{\tau\downarrow}}+\braket{\psi^\dag_{R\uparrow}\psi_{L\downarrow}}+\braket{\psi^\dag_{L\uparrow}\psi_{R\downarrow}}\right),
\end{equation}
where, by our MF ansatz, only the last two terms do not vanish. Employing Eq. (\ref{psi2C}), these yield  
\begin{equation}
    \braket{S^+} =\sum_{\mathbf k}\braket{C^\dag_{\mathbf k,R,\uparrow}C_{\mathbf k,L,\downarrow}}e^{-i\mathbf Q \cdot \mathbf r}+\sum_{\mathbf k}\braket{C^\dag_{\mathbf k,L,\uparrow}C_{\mathbf k,R,\downarrow}}e^{i\mathbf Q \cdot \mathbf r}=-\frac{1}{g_2}\left(\Delta^{+}e^{-i\mathbf Q \cdot \mathbf r}+(\Delta^{-})^*e^{i\mathbf Q \cdot \mathbf r} \right),
\end{equation}
in which the last equality follows from Eq. (\ref{OP}).
Assuming that the two independent components of the complex order parameter have equal magnitude but distinct arbitrary phases, i.e.  $\Delta^{(+)}=|\Delta|e^{i\alpha}$ and $\Delta^{(-)}=|\Delta|e^{i\beta}$, we get
\begin{equation}
\braket{S^+}=-\frac{|\Delta|}{g_2} e^{i \varphi}\cos(\mathbf Q \cdot \mathbf r+\phi)\, ;
\label{S+ExpVal}
\end{equation}
here $\phi$ and $\varphi$ are, respectively, the average and relative phases $\frac{1}{2}(\alpha \pm \beta)$.
This outcome can be interpreted as follows: by rotating the $x,y$ axes such that $x'$ axis aligns with the direction of $\mathbf Q$, the components of $\mathbf r$ in the newly defined axes, $x'$ and $y'$, are related through the transformation equation:
\begin{equation}
    \begin{pmatrix}
        x' \\ y'
    \end{pmatrix}=\begin{pmatrix}
        \cos(\theta) & \sin(\theta) \\ -\sin(\theta) & \cos(\theta)
    \end{pmatrix}\begin{pmatrix}
        x \\ y
    \end{pmatrix}
\end{equation}
where $\theta$ is the angle of $\mathbf Q$ with the $x$ axis. 
This rotation simplifies the expectation value of $S^+$ to a one-dimensional (striped) form, meaning that
\begin{equation}
    f(x,y) =\cos(Q_xx+Q_yy+\phi)\to f(x',y')=\cos(|Q|x'+\phi) \; .
\end{equation}

To obtain a pattern of the spin components in the spatial $x-y$ plane, we next extract the real and imaginary parts of Eq. (\ref{S+ExpVal}).  
This yields
\begin{equation}\label{Eq:TotalSx}
    \braket{S_{x}}=-\frac{4|\Delta|}{g_{2}}\cos\varphi\cos\left(\mathbf{Q}\cdot\mathbf{r}+\phi\right)\; ,
\end{equation}
\begin{equation}\label{Eq:TotalSy}
    \begin{split}
        \braket{S_{y}}=-\frac{4|\Delta|}{g_{2}}\sin\varphi\cos\left(\mathbf{Q}\cdot\mathbf{r}+\phi\right)\; .
    \end{split}
\end{equation}
A simple way to understand this result is by examining the local spin density vector of each helicity channel:
\begin{equation}\label{Eq:SSDW}
    \mathbf{S}^\tau(\mathbf{r})=A\left[\mathbf{e}_x \cos\right( \tau\mathbf{Q}\cdot\mathbf{r}+\theta_\tau\left)+\mathbf{e}_y \sin\right( \tau\mathbf{Q}\cdot\mathbf{r}+\theta_\tau\left) \right] \; ,
\end{equation}
where $\theta_+=\alpha$ and $\theta_-=-\beta$. Both channels represent a spiral spin density wave (SSDW) \cite{giamarchi2003quantum}, in which the spin polarization varies in direction but remains constant in magnitude.   The two local spin density operators, $\mathbf{S}^\tau(\mathbf{r})$, rotate about the spin quantization axis $\mathbf{e}_z$, with one rotating clockwise and the other counterclockwise. Since both share the same amplitude, the sum results in an antiferromagnetic order rather than spiraling spins, as depicted in Fig. \ref{fig:realSpaceSDW} of the main text.
\end{widetext}

\end{document}